\documentclass[twocolumn]{aastex63}

\usepackage{amsmath}
\usepackage{threeparttable}
\usepackage{multirow}

\newcommand{\bs}{\boldsymbol}

\newcommand{\akira}[1]{\textcolor{blue}{#1}}

\received{September 4, 2024}
\revised{March 25, 2025} 
\accepted{April 23, 2025}

\submitjournal{ApJ}

\shorttitle{An extended closure by LightGBM}
\shortauthors{Takahashi et al.}

\graphicspath{{./}{figures/}}

\begin{document}

\title{An Extended Closure Relation by LightGBM for Neutrino Radiation Transport \\ in Core-collapse Supernovae}

\email{tkst1228@g.ecc.u-tokyo.ac.jp}

\author{Shota Takahashi}
\affiliation{Department of Physics, The University of Tokyo, 7-3-1 Hongo, Bunkyo, Tokyo 113-0033, Japan}

\author[0000-0003-1409-0695]{Akira Harada}
\affiliation{National Institute of Technology, Ibaraki College, Hitachinaka 312-8508, Japan}
\affiliation{Interdisciplinary Theoretical and Mathematical Sciences Program (iTHEMS), RIKEN, Wako, Saitama 351-0198, Japan}

\author{Shoichi Yamada}
\affiliation{Advanced Research Institute for Science and Engineering, Waseda University, 3-4-1 Okubo, Shinjuku, Tokyo 169-8555, Japan}

\begin{abstract}
We developed a machine learning model using LightGBM, one of the most popular gradient-boosting decision tree methods these days, to predict the Eddington tensor, or the second-order angular moment, for neutrino radiation transport in core-collapse supernova simulations. We use not only the zeroth and first moments of the neutrino distribution function in momentum space as in ordinary closure relations but also information on the background matter configuration extensively. For training the model, we utilize some post-bounce snapshots from one of our previous Boltzmann radiation-hydrodynamics simulations; the Eddington tensor as well as the zeroth and first angular moments are calculated from the neutrino distribution function obtained in the simulation. LightGBM is light indeed, and its high efficiency in training enables us to feed a large number of features and figure out which features are more important than others. In this paper, we report the results of this feature engineering in addition to those of the training, validation, and generalization of our model. We find that the flux factor and non-local features are among the most relevant features; our LightGBM model can reproduce the Eddington factor better in general than the M1 closure relation, one of the most commonly employed algebraic closure relations at present; the generalization performance is also much improved from our previous model based on the deep neural network.

\end{abstract}


\section{Introduction}
Core-collapse supernovae (CCSNe) are explosive demises of massive stars. Despite extensive investigations, the detailed mechanism driving this explosion remains to be established. The neutrino heating \citep{Janka_2012, yamada2024review} is the mechanism that most researchers in society believe is most likely at present. The process proceeds as follows: the gravitational collapse of a massive star is halted when the core reaches nuclear density at the center. Then a shock wave is formed and sweeps the accreting outer core, photodissociating heavy nuclei therein. The shock energy is depleted through this endothermic reaction, which eventually leads to a shock stall. In the meantime, a proto-neutron star (PNS) is formed at the center and starts to emit neutrinos copiously. These neutrinos heat up matter behind the shock, re-energize the stagnant shock wave eventually, and revive it. It should be obvious that a good understanding of neutrino transport is imperative to establish the theory of the CCSN mechanism.

Numerical simulations have been extensively employed for this purpose \citep{Nagakura2019BoltzmannHydro, mezzacappa2020physical, Bollig_2021, Bruenn_2023, shibagaki2023threedimensional, wang2024supernova}. They comprise the computations of hydrodynamics, gravity, and neutrino transport and reactions. The neutrino transport, the target of this study, is described by the Boltzmann equation. Its faithful solution, although preferable, is numerically demanding because of its high dimensionality. The majority of CCSN simulations these days resort to some approximations. Among them, the truncated moment method is the most widely used \citep{Thorne1981RelativisticRT, shibata2011truncated, Cardall_2013, O_Connor_2015}. Instead of treating the distribution function in momentum space at each spatial point, it considers the angular moments thereof by integrating the Boltzmann equation with respect to the two angles that specify the neutrino momentum, thereby reducing the dimension.

The resultant equations for the moments of different orders are coupled with one another and form an infinite hierarchy. In numerical simulations, it is truncated at some order, typically the first-order. The efficacy of this truncated moment method hinges on the accuracy of the closure relation, which one imposes by hand to obtain unsolved moments from the solved ones. If one truncates the hierarchy at the first-order, the Eddington tensor, or the second moment of the distribution function, is normally given by the closure relation as a function of the zeroth and first-order moments. Among various closure relations proposed in the literature to date, the M1 closure relation is the most popular these days \citep{Levermore1984}. According to the comparison of the Eddington tensors derived by the direct integration of the distribution functions obtained in the CCSN simulations that solve the Boltzmann equation faithfully, which we refer to as the Boltzmann simulations hereafter, and the Eddington tensors given by the M1 closure relation for the same data, the diagonal and off-diagonal components of the Eddington tensor may develop discrepancies of several tens of percent and a factor of 2, respectively \citep{2019ApJ...872..181H, Iwakami_2022}.

Some efforts are still being made to enhance the accuracy of closure relations. \cite{2021PhRvD.103l3025N} suggested a fitting formula for the Eddington factor, the largest eigenvalue of the Eddington tensor, based on the spherically symmetric Boltzmann simulations. Though it offers a fairly good formula, more improvement is desirable to apply it to the multi-dimensional simulations. In a separate endeavor, \citet{harada2022deep} developed a tensor basis neural network (TBNN), a deep neural network that provides the Eddington tensor from the neutrino energy density and flux, i.e., the zeroth and first-order moments and the local matter velocity. The network was trained by using the data of a Boltzmann simulation. Although the TBNN-closure outperforms the M1 closure in terms of accuracy, there remains room for improvement both in the accuracy and in the generalization performance.

This paper aims to improve the closure relation by machine learning techniques. From our experience in the previous attempt with the TBNN \citep{2019ApJ...872..181H}, we think that a substantial improvement will be achieved only by providing more information other than the two lowest order moments to the machine learning model. In fact, \citet{2019ApJ...872..181H} found that the information on local matter motions was useful as an input to the TBNN. One particularly powerful technique to identify relevant features is the decision tree for regression problems (see Section 2.2 for details). We adopt Light Gradient Boosting Machine (LightGBM) in this paper. It is known that LightGBM can handle a large number of features very efficiently. As one of the decision tree algorithms, on the other hand, it may lack sufficient smoothness in expressing functions. We attach more importance to the former capability in this study.

This paper is organized as follows. In section 2, we briefly review the M1 closure relation and LightGBM, the machine learning model employed in this paper. Section 3 provides the details of the training: data, feature engineering, validation, and hyperparameters. Then, in section 4, we present the main results on the performance of our novel models making comparisons with the Eddington tensors calculated directly from the original data obtained by our Boltzmann simulation as well as with the Eddington tensors derived from the M1 closure relation applied to the zeroth and first angular moments calculated from the same original data. Finally, section 5 is the conclusion. Throughout this paper, we employ the units of $c=1$, where $c$ is the light speed. The metric signature is $(-+++)$. Greek and Latin indices run over $0$--$3$ (spacetime) and $1$--$3$ (spatial).

\section{Overviews of the M1 closure relation and LightGBM}
The purpose of this paper is to improve by machine learning techniques the estimation of the Eddington tensor in the truncated moment method for the neutrino transport in CCSN simulations. This section is devoted to a brief review of the M1 closure relation in section \ref{sec:closure} and LightGBM in section \ref{sec:lightgbm}.

\subsection{M1 closure relation}
\label{sec:closure}
The neutrino population in CCSNe is described by the distribution function $f(x,p)$ in phase space, in which $x$ and $p$ are the spacetime coordinates and momentum of neutrinos. It is governed by the Boltzmann equation. The high dimensionality and stiff source terms of this equation pose numerical challenges with current computational resources. Because of its high numerical cost, most researchers are opting for the truncated moment method nowadays.

Since the formulation of the moment method is described in \citet{Thorne1981RelativisticRT, shibata2011truncated} in detail, we give only its outline in the following. We first define the unprojected second moment of the distribution function as\footnote{We follow the definition in \citet{2018ApJ...854..136N,2019ApJ...872..181H,2020ApJ...902..150H}, which is slightly different from that in \citet{shibata2011truncated}.}
\begin{equation}
    M^{\alpha \beta}\left(\varepsilon, x \right) = \int f\left(x, p^{\prime} \right) \delta\left(\frac{\varepsilon^3}{3}-\frac{\varepsilon^{\prime 3}}{3}\right) p^{\prime \alpha} p^{\prime \beta} d V_{p^\prime},
\end{equation}
where $\varepsilon^{\prime}=-u_\mu p^{\prime \mu}$ is the neutrino energy measured in the fluid-rest (FR) frame with $u$ being the fluid four-velocity. By applying appropriate projections to this moment, we can derive the energy densities $E$, energy flux $F^i$, and the stress tensors $P^{ij}$ in both the FR and laboratory (LB) frames:
\begin{equation}
    E_{\rm FR} = M^{\alpha\beta} u_\alpha u_\beta,\;\;E_{\rm LB} = M^{\alpha\beta} n_\alpha n_\beta,
\end{equation}
\begin{equation}
    F_{\rm FR}^i = -M^{\alpha\beta} u_\alpha h_\beta{}^i,\;\; F_{\rm LB}^i = -M^{\alpha\beta} n_\alpha \gamma_\beta{}^i,
\end{equation}
and
\begin{equation}
    P_{\rm FR}^{ij} = M^{\alpha\beta} h_\alpha{}^i h_\beta{}^j,\;\; P_{\rm LB}^{ij} = M^{\alpha\beta} \gamma_\alpha{}^i \gamma_\beta{}^j,
\end{equation}
where $n$ is the unit normal vector to the spatial hypersurface with $t = \rm{const.}$;  $\gamma_\alpha{}^i = \delta_\alpha{}^i + n_\alpha n^i$ is the projector onto that surface with $\delta_{\alpha}{}^\beta$ being the Kronecker's delta, whereas $h_\alpha{}^i = \delta_\alpha{}^i + u_\alpha u^i$ is the projector onto the plane perpendicular to $u^{\alpha}$. In the truncated moment formalism, the time evolutions of $E_{\rm LB}$ and $F_{\rm LB}^i$ are solved normally by setting a closure relation to give $P_{\rm LB}^{ij}$ as a function of $E_{\rm LB}$ and $F_{\rm LB}^i$\footnote{In order to solve the evolution of spectrum in general relativity, the third moment of the distribution function is also required. We ignore it in this paper for simplicity, though.}. These three quantities are essentially the lowest three angular moments. Their exact relations are given, e.g., in \citet{shibata2011truncated}.

The closure relation is normally imposed to the Eddington tensor $k^{ij}$ defined as the stress tensor divided by the energy density. A common way to construct it is to employ the interpolation
of the optically thick and thin limits:
\begin{equation}
    P_{\rm LB}^{ij} = \frac{3p_\nu-1}{2} P_{\rm thin}^{i j} + \frac{3(1-p_\nu)}{2} P_{\rm thick}^{i j}.
\end{equation}
In the above expression, $P^{ij}_{\rm thin}$ and $P^{ij}_{\rm thick}$ are the stress tensors in the optically thick and thin limits, resepctively, and can be expressed with the energy density and energy flux density as given shortly; $p_{\nu}$ is the Eddington factor and should be also prescribed in the truncated moment method. Then the Eddington tensor is obtained as
\begin{equation}
    k^{ij} = \frac{P_{\rm LB}^{ij}}{E_{\rm LB}}.
\end{equation}
The functional forms of the stress tensor in the two limits and of the Eddington factor is given as follows \citep{shibata2011truncated}. In the optically thin limit, neutrinos move freely in one direction specified by the flux vector. The stress tensor $P_{\rm thin}^{ij}$ is then given in the LB frame as
\begin{equation}
    P_{\rm thin}^{i j} = E_{\rm LB} \frac{F_{\rm LB}^i F_{\rm LB}^j}{|F_{\rm LB}|^2}.
\end{equation}
In the optically thick limit, on the other hand, neutrinos are in thermal equilibrium with matter, and are isotropic in the FR frame. In the LB frame, $P_{\rm thick}^{ij}$ is given by the Lorentz transformation as
\begin{eqnarray}
    P_{\rm thick}^{i j} &=& \frac{1}{3} E_{\rm FR}(\gamma^{ij} + 4 v^i v^j) \nonumber \\
    & &  + F_{\rm FR}^i v^j +  F_{\rm FR}^j v^i.
\end{eqnarray}
As for the Eddington factor, the M1 closure scheme adopts the following expression proposed by \cite{Levermore1984}:
\begin{equation}
    p_\nu = \frac{3 + 4 \chi^2}{5 + 2 \sqrt{4 - 3 \chi^2}},
\end{equation}
where $\chi=|F_{\rm{FR}}|/E_{\rm{FR}}$ is referred to as the flux factor.

\subsection{Light Gradient Boosting Machine}
\label{sec:lightgbm}
The decision tree (DT) is a machine learning algorithm that performs classification or regression tasks by recursively partitioning a dataset based on certain criteria \citep{pml1Book}. Suppose that we find a functional relation between an $n$-dimensional input vector $\bs x$ and an output real number\footnote{If the output is integer or discrete number, the task is a classification.} $y$ from a dataset of size $N$: $\{(\bs x^i, y^i)\}_{i=1}^{N}$. In the DT method, each data $(\boldsymbol x^j, y^j)$ is binary-classified according to $\boldsymbol{x}^j$, and the two destinations are called nodes. The data classified to a certain node is further classified to one of its two daughter nodes recursively. This process hence forms a tree graph. In the DT regression, each node has the prediction value $\hat{y}_k = \sum_{j\in D_k} y^j/|D_k|$, where $D_k$ is a subset of the data classified to node $k$ and $|D_k|$ is its size. The classification criterion at each node is determined so that it would minimize the mean squared error of the predicted value $\hat{y}_{R(L)k}$ at the right (left) daughter node of node $k$. This binary classification is repeated until the error thus obtained gets small enough. Then the output of the DT is the prediction value $\hat{y}^j$ of the node, to which the input $\boldsymbol{x}^j$ is finally classified.

The Gradient Boosting DT \citep[GBDT,][]{Friedman1999GreedyFA} is an improved version of DT. The output of a single DT for regression has large errors normally. We can improve a prediction by training another DT for these errors. With recursive corrections with many DTs, the GBDT achieves an accurate prediction. Suppose that $f_\ell (\boldsymbol{x})$ is the output of the $\ell$-th DT for the input vector $\boldsymbol{x}$. In the least-square GBDT, each DT is trained so that the following functions
\begin{equation}
    \sum_{j=1}^N\left\{\left(y^j-\eta\sum_{m=0}^{\ell-1}f_m(\boldsymbol{x}^j)\right)-f_\ell(\boldsymbol{x}^j)\right\}^2
\end{equation}
should be minimized from $\ell=1$. In the above expression, $0< \eta < 1$ is the learning rate employed to avoid overfitting and $f_0(\boldsymbol{x})$ is given conventionally by $\sum_{j=1}^N y^j/N$. Then, the prediction of GBDT is given by $\hat{y}^j = \eta \sum_{\ell=0}^M f_\ell (\boldsymbol{x}^j)$, where $M$ is the total number of DTs.

\begin{figure}
    \centering
    \includegraphics[width=1\linewidth]{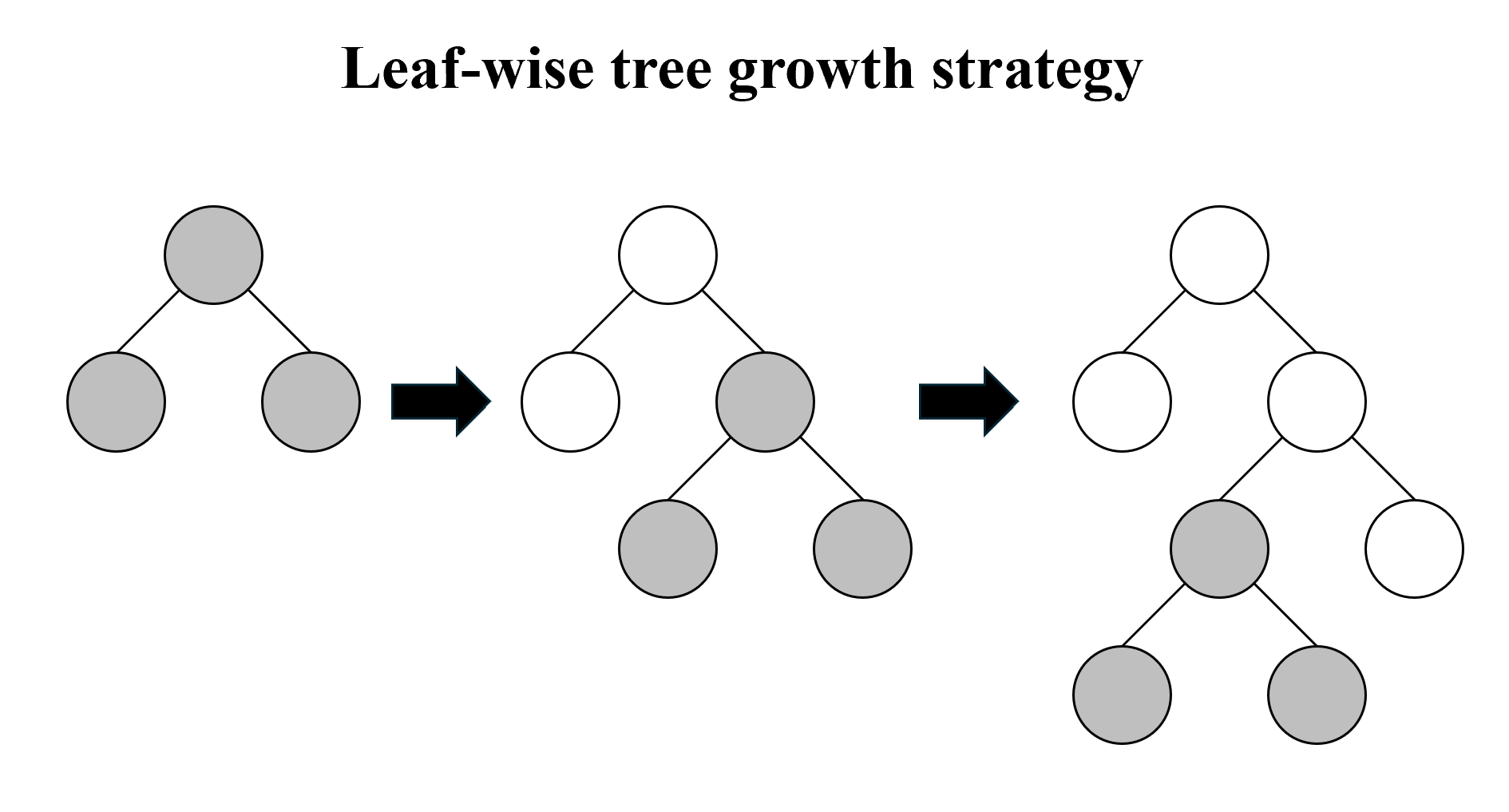}
    \caption{A 
    schematic diagram explaining the process of the leaf-wise tree growth strategy. Each node represents a leaf in the decision tree, with the grey nodes representing the leaves that are split at each step.}
    \label{fig:leaf_wise_tree}
\end{figure}

LightGBM (Light Gradient Boosting Machine) is a
variant of GBDT introduced by \cite{NIPS2017_6449f44a}. LightGBM incorporates four techniques to enhance GBDT:  (1) leaf-wise tree growth that prioritizes creations of daughter and granddaughter nodes than those of cousin nodes, suppressing the growth of 
the nodes that focus on irrelevant variables for regression, as shown in figure \ref{fig:leaf_wise_tree}, (2) a histogram-based algorithm, which reduces numerical cost by binning data, (3) the gradient-based one-sided sampling (GOSS) to reduce the training data by focusing on the critical elements, and (4) the exclusive feature bundling (EFB), which combines features that are not zero simultaneously. Thanks to these techniques, LightGBM excels in computational and memory efficiency in the training, compared to traditional models. It can hence accelerate the search for relevant variables in the closure relation.

\section{Method}
This section provides a detailed description of the model developed in this paper. Section 3.1 concerns the data used in this study, section 3.2 discusses the feature engineering, and section 3.3 explains the model.

\subsection{Data Description}
The data set used in this study is composed of five snapshots taken from a 2D axisymmetric CCSN simulation performed with the Boltzmann-radiation-hydrodynamics code \citep{SumiyoshiYamada2012, Nagakura2014ThreeDimensional, Nagakura2019BoltzmannHydro, Nagakura2019ProtoNeutronStars}. This code solves the Boltzmann equation for neutrinos, the hydrodynamics equations for stellar gas, and the Poisson equation for gravity simultaneously. We adopt Furusawa--Togashi EOS \citep{2017JPhG...44i4001F}, which is based on the variational method and is extended to handle a large number of nuclei in the nuclear statistical equilibrium. The progenitor is a non-rotating star with the ZAMS mass of $15\,M_{\odot}$ modeled by \cite{RevModPhys.74.1015}. For a detailed description of this code we refer readers to \cite{SumiyoshiYamada2012}, \cite{Nagakura2014ThreeDimensional}, \cite{Nagakura2019BoltzmannHydro}, and \cite{Nagakura2019ProtoNeutronStars}. Whereas we consider three neutrino species: electron-type neutrinos and anti-neutrinos as well as heavy-lepton-type neutrinos, we employ the results for the electron-type neutrino alone in this study. The grid sizes for radius $N_r$, zenith angle $N_\theta$, energy $N_\varepsilon$, momentum angles $N_{\theta_\nu}$, $N_{\phi_\nu}$ are $(N_r, N_\theta, N_\varepsilon, N_{{\theta}_{\nu}}, N_{{\phi}_{\nu}}) = (384, 128, 20, 10, 6)$. The computational zone covers the region up to 5000 km from the stellar center, with $0<\theta \rm{(rad)}<\pi$ and $0<\varepsilon \rm{(MeV)}<300$. In this study, on the other hand, we utilize the data within the ranges of $0<r\rm{(km)}<200$ because this is the region of the greatest relevance for the neutrino heating and shock revival. We extract the distribution function of electron-type neutrino in this region from the simulation data at the post-bounce time of $300\,{\rm ms}$ and use it as the training data. For validation, we employ four other distribution functions sampled at $t=100\,\rm{ms}$, $t=150\,\rm{ms}$, $t=200\,\rm{ms}$ and $t=250\,\rm{ms}$ after bounce. Note that the CCSN simulation we adopt in this paper is identical to what was used in \cite{harada2022deep}, but the radial extent and the number of snapshots are expanded.

We will construct a machine learning model to give the Eddington tensor as a function of some variables, or features, that include the lower moments. Those we refer to as the basic features \replaced{are listed in Table \ref{tab:basic_feat}.}{includes the variables to construct the conventional closure relations.}
As discussed later, additional features are generated from these basic ones through feature engineering. The basic and additional features are collectively treated as inputs to the machine.

The output of the machine is the Eddington tensor. In this paper, we do not project it onto the local orthonormal vectors: $e_r$, $e_\theta$, and $e_\phi$, of the spherical coordinates in the LB frame. Instead, we employ the orthonormal vectors, $e_{{\rm FP}, i}$, one of which is aligned with the neutrino flux, and the other two are orthogonal to it and also to each other. We refer to the 3-dimensional frame defined by these orthonormal vectors as the flux-projected (FP) frame, and the Eddington tensor represented in this frame is denoted as $k^{ij}_{\rm FP}$ hereafter. The FP frame is constructed for each neutrino energy as follows: $e_{{\rm FP}, 1}$ is the unit vector parallel to the neutrino flux. The choice of the other two unit vectors is rather arbitrary. In this paper we obtain $e_{{\rm FP}, 2}$ from $e_{\theta}$ by subtracting the component parallel to $e_{{\rm FP}, 1}$, i.e., via the Gram-Schmidt orthogonalization. As a result, $e_{{\rm FP}, 2}$ so obtained lies in the plane spanned by the flux and $e_{\theta}$. The last vector is derived so that $e_{{\rm FP},1}$, $e_{{\rm FP}, 2}$ and $e_{{\rm FP},3}$ should form a right-handed orthonormal system. The Eddington tensor in the FP frame is then related to that in the LB frame as $k_{{\rm FP}}^{ij} e_{{\rm FP},i}^{l} e_{{\rm FP},j}^{m} = k_{\rm LB}^{lm}$, where the superscript of $e_{{\rm FP}}$ specifies the components and the repeated indices are summed over.

The reason why we employ the FP frame is the following. The Eddington tensor is not ``aligned'' in general with the coordinates we choose rather arbitrarily. Since the Eddington tensor is a symmetric second-rank tensor, it becomes diagonal in the appropriately chosen orthonormal frame, which is not equal to the LB frame in general. Suppose that the neutrino distribution in momentum space is axisymmetric with respect to a certain direction misaligned with any one of the orthonormal vectors in the LB frame. Instead, the symmetry axis is parallel to the flux vector in this case. In fact, the Eddington tensor is diagonal in the FP frame. It should be stressed that the neutrino distribution in momentum space is not axisymmetric in general, except for the notable case of spherical symmetry. For example, the neutrino flux from a rotating oblate PNS in space is non-radial and has even nonvanishing $\phi$-component, since neutrinos carry non-zero angular momenta. Then the Eddington tensor in the LB frame has substantial off-diagonal components \citep{2019ApJ...872..181H,Iwakami_2022}. We expect, on the other hand, that the off-diagonal components will be smaller in the FP frame even in this case.


\subsection{Feature Engineering}
The closure relation is originally meant to give high-order moments such as the Eddington tensor in terms of lower-order moments like the energy density and flux. In our previous attempt to build a deep neural network to do this task, we realized that the lower-order moments are just not sufficient and added some other local features such as the matter velocity and its shear to the input. In this paper, we extend this idea. Thanks to the efficiency of LightGBM, we are able to incorporate much more information, some of which are non-local. It is true that the more model-specific information we employ, the less general the machine becomes, but the purpose of this paper is to construct a machine learning model that we can apply to the CCSN simulation with the truncated moment method. 

In providing a large number of features to the machine, we find feature engineering very important. By feature engineering, we mean finding better combinations of the original features, normalizing features, and, if necessary, removing the basic features after using them for feature engineering. In principle, this process does not increase the information included in the original features. If the output depends on a function of rather involved combinations of these features, the machine may need a long time or even fail to train itself even if those features contain sufficient information. On the other hand, if the human knows a priori those combinations of the original features and teaches them to the machine, we can expect the training efficiency to be improved substantially. Unfortunately, we do not know what the best combinations should be. Thanks to the efficiency of LightGBM, we can try various combinations. The engineered features given below are indeed obtained that way.

As mentioned above, we will incorporate non-local quantities to the input features. This is probably understandable if one recalls the fact that the radiative transfer is most difficult to treat in the semi-transparent region, where the mean free path of neutrinos becomes comparable to the scale height of matter, and the neutrino distribution is indeed not determined locally. \deleted{It is noted that the basic features in Table 1 include the optical depth, which is non-local information. As explained below, we add more.}

\added{The features employed as inputs in this paper are summarized in Table \ref{tab:features}. The basic features (ll. 1--6 of Table \ref{tab:features}) are those utilized in the conventional closure relations and some other variables encoding hydrodynamical structure.} The engineered features thus produced are \added{further} classified into three types: \replaced{advanced features}{combined features}, \replaced{spatial-shift}{non-local} features, and \replaced{spatial-difference features}{differential features}. \added{
The combined features are produced from the basic ones to capture their complex expressions and/or specific physical properties. The non-local features incorporate information from neighboring grid points to account for possible spatial correlations beyond the local grid. The differential features are spatial gradients, emphasizing changes in quantities in the radial or angular direction. Their detailed definitions and roles will be explained later. These three types of engineered features collectively enhance the model by capturing both local and global characteristics of the neutrino radiation field.}

\akira{\begin{table*}[h]
\centering
\caption{The features employed in the model}
\begin{threeparttable}
\begin{tabular}{ccccc}
\hline
Class & Feature & Definition & Dimension & Adopted\tnote{a}\\
\hline
\multirow{6}{*}{basic}&Neutrino energy & $\varepsilon$ & $\rm{(MeV)}$ & D/OD\\
&Optical depth & $\tau$ & - & OD\\
&Baryon mass density & $\rho_{\text{baryon}}$ & $\rm{(g/cm^3)}$ & OD\\
&Energy density (LB) & $E_{\text{LB}}$& $\rm{(erg/cm^3)}$ & OD\\
&Flux (LB) & $F_{\text{LB}}^i\,(i=r,\theta,\phi)$& $\rm{(erg/cm^3)}$ & OD\\
&Fluid velocity (LB) & $v^i\,(i=r,\theta,\phi)$& $\rm{(cm/s)}$ & OD\\
\hline
\multirow{11}{*}{combined}&Flux factor (LB) & $\chi_{\rm LB}=|F_{\rm LB}|/E_{\rm LB}$ & - & D/OD\\
&collinearity parameter &$F_{\rm{LB}}^k v_k/E_{\rm{LB}}|v|$ & - & D/OD\\
&Sign of $v^{j}\;(j=1,2)$ (FP) & $I(v_{\text{FP}}^j \geq 0)$\tnote{b} $\;(j=1,2)$ & - & D/OD \\
&In neutrinosphere & $I(\tau \geq 5)$ & - & D/OD\\
&Logarithm of flux factor & $\log(\chi_{\text{LB}})$ & - & D/OD\\
&Logarithm of optical depth & $\log(\tau)$ & - & D/OD\\
&Logarithm of baryon mass density & $\log(\rho_{\text{baryon}}/({\rm g/cm^3}))$ & - & D/OD\\
&Fluid speed & $|v|$ & (cm/s) & OD\\
&Fluid velocity (FP) & $v_{\rm FP}^i\;(i=1,2,3)$ & (cm/s) & OD\\
&Normalized fluid velocity (FP) & $v_{\rm FP}^i/|v|\;(i=1,2,3)$ & - & OD\\
&Neutrino mean velocity (FP) & $F_{\rm FP}^i/E\;(i=1,2,3)$ & - & OD\\
\hline
\multirow{4}{*}{near non-local} & Flux factor (LB) & $\chi_{\rm LB}=|F_{\rm LB}|/E_{\rm LB}$ & - & D/OD\\
&collinearity parameter &$F_{\rm{LB}}^k v_k/E_{\rm{LB}}|v|$ & - & D/OD\\
& Fluid velocity (FP) & $v_{\rm FP}^i\;(i=1,2)$ & (cm/s) & OD \\
& Normalized fluid velocity (FP) & $v_{\rm FP}^i/|v|\;(i=1,2)$ & - & OD\\
&Fluid speed & $|v|$ & (cm/s) & OD\\
&Neutrino mean velocity (FP) & $F_{\rm FP}^i/E\;(i=1,2)$ & - & OD\\
\hline
\multirow{2}{*}{far non-local} & Flux factor (LB) & $\chi_{\rm LB}=|F_{\rm LB}|/E_{\rm LB}$ & - & D/OD\\
&collinearity parameter &$F_{\rm{LB}}^k v_k/E_{\rm{LB}}|v|$ & - & D/OD\\
\hline
\multirow{6}{*}{differential} & Flux factor (LB) & $\partial_j \chi_{\rm LB}\;(j=r,\theta)$ & ($\rm cm^{-1}$) or - & D/OD\\
& collinearity parameter &$\partial_j (F_{\rm{LB}}^k v_k/E_{\rm{LB}}|v|)\;(j=r,\theta)$ & ($\rm cm^{-1}$) or - & D/OD\\
& Fluid velocity (FP) & $\partial_j v_{\rm FP}^i\;(i=1,2;\;j=r,\theta)$ & ($\rm s^{-1}$) or ($\rm cm/s$) & OD \\
& Normalized fluid velocity (FP) & $\partial_j (v_{\rm FP}^i/|v|)\;(i=1,2;\;j=r,\theta)$ & ($\rm cm^{-1}$) or - & OD\\
&Fluid speed & $\partial_j |v|\;(j=r,\theta)$ & ($\rm s^{-1}$) or ($\rm cm/s$) & OD\\
&Neutrino mean velocity (FP) & $\partial_j (F_{\rm FP}^i/E)\;(i=1,2;\;j=r,\theta)$ & ($\rm cm^{-1}$) or - & OD\\
\hline
\end{tabular}
\begin{tablenotes}
\item[a] ``Adopted'' column indicates whether each feature is adopted in both diagonal and off-diagonal prediction (O/OD) or in only off-diagonal prediction (OD).
\item[b] $I$ is the indicator function, which equals 1 if the condition inside the parentheses is true, and 0 otherwise.
\end{tablenotes}
\end{threeparttable}
\label{tab:features}
\end{table*}}

\replaced{Listed in Table \ref{tab:Advanced_features} are the combined features considered here.}{The combined features considered here is listed as ll. 7--17 in Table \ref{tab:features}.}
Particularly important for the improvement of the model are the flux factor in the LB frame $\chi_{\rm LB} = |F_{\rm LB}^i|/E_{\rm LB}$, the collinearity parameter $\frac{F_{\rm{LB}}^k v_k}{E_{\rm{LB}} |v^i|}$ that measures the alignment of the neutrino flux and matter velocity, and the parameter we call the ''neutrinosphere indicator'' which is a binary indicator of whether the point is well inside the neutrinosphere or not, i.e., it is 1 if the optical depth $\tau$ is greater than 5 and 0 otherwise.

\added{
The flux factor plays a particularly important role among these quantities. The flux factor reflects the forward-peakedness of the neutrino angular distribution in momentum space \citep{Levermore1984}. Since it is a dimensionless quantity, it is expected to exhibit smaller variations across the snapshots at different times compared to other physical quantities (e.g., energy density or the flux), thereby enhancing the generalization performance itself.}

The \replaced{spatial-shift}{non-local} features are the features concerning the neighboring grid points. The closure relation is normally local; i.e., it is a relation among physical quantities at the same location. It turns out, however, that the inclusion of the neighboring physical quantities as input improves the accuracy of the inference. More concretely, we incorporate
\replaced{the flux factor $\chi_{\rm LB}=|F_{\rm LB}^i|/E_{\rm LB}$ and the collinearity parameter $\frac{F_{\rm{LB}}^jv_j}{E_{\rm{LB}}|v^i|}$ not only on the grid point of concern but also on the neighboring grid points as follows. Let $Q_{k,l}$ represent one of the aforementioned quantities at the $k$-th radial grid point and the $l$-th angular grid point; we then incorporate $Q_{k-5i, l}$ (where $i=-1, 1, 2, \dots, 6$), $Q_{k+1, l}$, $Q_{k-1, l}$, $Q_{k, l+1}$ and $Q_{k, l-1}$ as components of the input feature vector for $Q_{k,l}$.}{the features listed as ll. 18--23 in Table \ref{tab:features} not only on the grid point of concern but also on the neighboring grid points. Here, we further classify the non-local features into near non-local and far non-local ones. Let $Q_{k,l}$ represent one of the listed features at the $k$-th radial grid point and the $l$-th angular grid point; we then incorporate $Q_{k-1,l}$, $Q_{k+1,l}$, $Q_{k,l-1}$, and $Q_{k,l+1}$ as near non-local features (ll. 18--21 in the table), and $Q_{k-1,l}$, $Q_{k+1,l}$, $Q_{k,l-1}$, $Q_{k,l+1}$, and $Q_{k-5i, l}$ ($i=-1, 1, 2, \dots, 6$) as far non-local features (ll. 22--23 in the table)}\footnote{If it goes beyond the spatial boundaries, those points are simply ignored.}. We find in particular that the incorporation of the inner points, $Q_{k-5i, l}$ (where $i=1, 2, \dots, 6$) and $Q_{k-1, l}$, significantly improves the prediction accuracy. The reasoning here is that since neutrinos predominantly move outward, the upstream information, i.e., the information on the radially inner points is particularly valuable.

We also find it useful to employ the derivatives of features \added{listed as ll. 24--29 in Table \ref{tab:features}}:
\begin{equation}
    \frac{\Delta Q_{k,\ell}}{\Delta r} = \frac{Q_{k,\ell}-Q_{k-1,\ell}}{r_{k,\ell}-r_{k-1,\ell}}.
\label{eq:diff_feat_radius}
\end{equation}
\begin{equation}
    \frac{\Delta Q_{k,\ell}}{\Delta \theta} = \frac{Q_{k,\ell}-Q_{k,\ell-1}}{r_{k, \ell}(\theta_{k,\ell}-\theta_{k,\ell-1})}.
\label{eq:diff_feat_theta}
\end{equation}
to which we refer as the \replaced{spatial-difference feature}{differential features}. Although it may seem redundant, we include both the \replaced{spatial-shift}{non-local} and \replaced{difference features}{differential features} in the input as we do not know which one is more effective. Note that irrelevant features are automatically ignored in LightGBM.

\added{The physical significance of the non-local and differential features may be understood as follows. The neutrino radiation field at each spatial point is a combination of the local emission from its close vicinity and the non-local contribution from the radiation of the PNS sitting at the center \citep{Harada_2019}. The former may be accounted for by the local features such as the baryon density and the collinearity parameter (which conveys the information on the local fluid motions), while the latter may be described only by the non-local and differential features. By the incorporation of the non-local information in the angular direction, we expect the inference to be further enhanced.}

\added{It is worth noting that different features were utilized for the diagonal and off-diagonal components in our LightGBM model. Though the all features listed in Table \ref{tab:features} are utilized in the off-diagonal component prediction, only a part of the list is employed for the diagonal component prediction. The total number of the features adopted in the diagonal component prediction is $1$ (basic) $+8$ (combined) $+2\times4$ (near non-local) $+2\times 7$ (far non-local) $+2\times 2$ (differential) $=35$, while that in off-diagonal prediction is $10$ (basic) $+18$ (combined) $+9\times 2$ (differential) $+9\times 4$ (near non-local) $+2\times 7$ (far non-local) $=96$. For example, features such as fluid velocity, which were removed when predicting diagonal components, are included in off-diagonal prediction. When predicting diagonal components, the prediction accuracy significantly decreased; however, in predicting off-diagonal components, the accuracy was not greatly compromised. Off-diagonal components are more significantly affected by factors such as fluid motion compared to diagonal components, which suggests that background information may be more important for their prediction.}

\subsection{Models}
\begin{figure}
    \centering
    \includegraphics[width=1\linewidth]{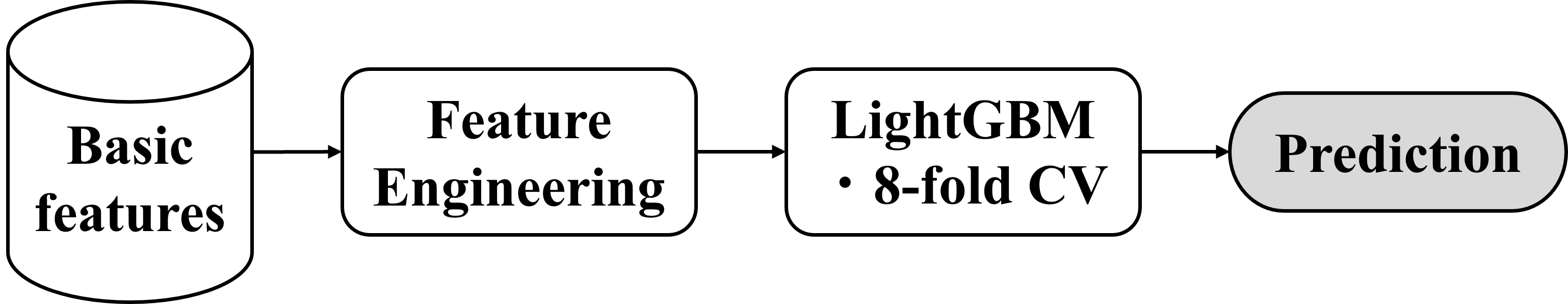}
    \caption{A pipeline of the model adopted in this paper. See the text for details.}
    \label{fig:pipeline}
\end{figure}

The pipeline of the model employed in this study is shown in Figure \ref{fig:pipeline}. After creating new features from the basic features through feature engineering, we conduct training with the 8-fold cross-validation (CV), in which the data set is randomly divided into 8 subsets to build 8 models. One of them serves as the validation set and the remaining 7 as the training data. They make predictions individually and the arithmetic mean of those predictions is adopted as the final prediction of this model. Such an approach is known to effectively prevent the model from over-fitting a particular dataset, thereby improving generalization.

The main hyperparameters and their values in the LightGBM model are listed in Table \ref{tab:hyperparams}. We employ the early stopping training, i.e., the training is terminated once the mean squared error is no longer lowered remarkably. It is also useful to avoid overfitting. As a consequence, we can set a sufficiently large value to \texttt{n\_estimators}, the number of decision trees to be built. Though it is not our main goal in this paper to fine-tune the hyperparameters, we set some hyperparameters separately for the diagonal components of the Eddington tensor and for the off-diagonal components.

\begin{table*}
\centering
\begin{tabular}{|l|l|l|}
\hline
Parameter & Description & Value\\ \hline
\texttt{method} & The method to be applied for training. &  \texttt{goss} \\ \hline
\texttt{device} & The device to be used for training. & \texttt{gpu} \\ \hline
\texttt{importance\_type} & Method for measuring the importance of features. & \texttt{gain} \\ \hline
\texttt{objective} & The objective function to be optimized. & \texttt{mse} \\ \hline
\texttt{n\_estimators} & The number of decision trees to be built. & 50000 \\ \hline
\texttt{num\_leaves} & The maximum number of leaves in one tree. & 256 \\ \hline
\texttt{learning\_rate} & The learning rate. & $8.75 \times 10^{-3}$ \\ \hline
\texttt{max\_depth} & The maximum depth of a tree. & 10 \\ \hline
\texttt{min\_child\_samples} & The minimum number of data points required in a leaf node. & 20 \\ \hline
\texttt{reg\_alpha} & Coefficient for L1 regularization. & $10^{-4}$ (diagonal) \\ & & $10^{-2}$ (off-diagonal) \\ \hline
\texttt{reg\_lambda} & Coefficient for L2 regularization. & $10^{-3}$ (diagonal) \\ & & $10^{-1}$ (off-diagonal) \\ \hline
\texttt{min\_split\_gain} & The minimum loss reduction required to make a split. & $1.25 \times 10^{-5}$ \\ \hline
\texttt{min\_child\_weight} & The minimum sum of instance weights (Hessian) in a leaf. & $10^{-4}$  \\ \hline
\texttt{feature\_fraction} & The fraction of features used for training each tree. & 0.9 \\ \hline
\end{tabular}
\caption{Key Hyperparameters and their Values adopted in LightGBM}
\label{tab:hyperparams}
\end{table*}

\section{Results}
\label{sec:results}
In this paper, the data extracted from the snapshot at 300 ms after bounce is used as the training set. The target variable is the Eddington tensor in the FP frame $k_{\rm{FP}}^{ij}$. We focus on the estimation of its 11-, 22-, 33-, and 12-components, since they are the diagonal and main off-diagonal elements, respectively. The test data were collected from the snapshots at 100 ms, 150 ms, 200 ms, and 250 ms after bounce. In the following, the diagonal and off-diagonal components are considered separately in turn.

\subsection{Diagonal Components}
The diagonal components of the Eddington tensor play the main role in neutrino transport, and hence their accurate estimation is crucially important. The conventional M1 closure relation does not reproduce it perfectly \citep{2018ApJ...854..136N,2019ApJ...872..181H,2020ApJ...902..150H,2022ApJ...933...91I}, and the machine-learning closure is expected to provide better predictions. 

\deleted{Figure \ref{fig:loss_plot} shows the learning curve in the training of the 11-component of the Eddington tensor.} The training is conducted to minimize the Mean Squared Error (MSE):
\begin{equation}
    L^{(2)} = \frac{1}{|D|} \sum_{D}|k_{\rm boltz, FP}^{ij}-k_{\rm infer, FP}^{ij}|^2,
\label{eq:loss_function}
\end{equation}
where the sum is taken over \deleted{the subset $D_\varepsilon$ of} the \added{whole} data \deleted{corresponding to a particular neutrino energy} and \replaced{$|D_\varepsilon|$}{$|D|$} is the number of its elements; $k_{\rm boltz, FP}^{ij}$ is the Eddington tensor obtained by the Boltzmann-radiation-hydrodynamics simulation whereas $k_{\rm infer, FP}^{ij}$ is the one obtained with our LightGBM model. \deleted{As the solid lines show, the training is successful with the values of MSE decreasing monotonically down to $<10^{-5}$ for all 8 models. These lines are terminated at different numbers of estimators because of the early stopping. The evaluation of the validation data also presented in the figure as dashed lines indicates that the overfitting is avoided reasonably well indeed. Similar results are obtained for other components of the Eddington tensor.}

\begin{figure*}
    \centering
    \includegraphics[width=1\linewidth]{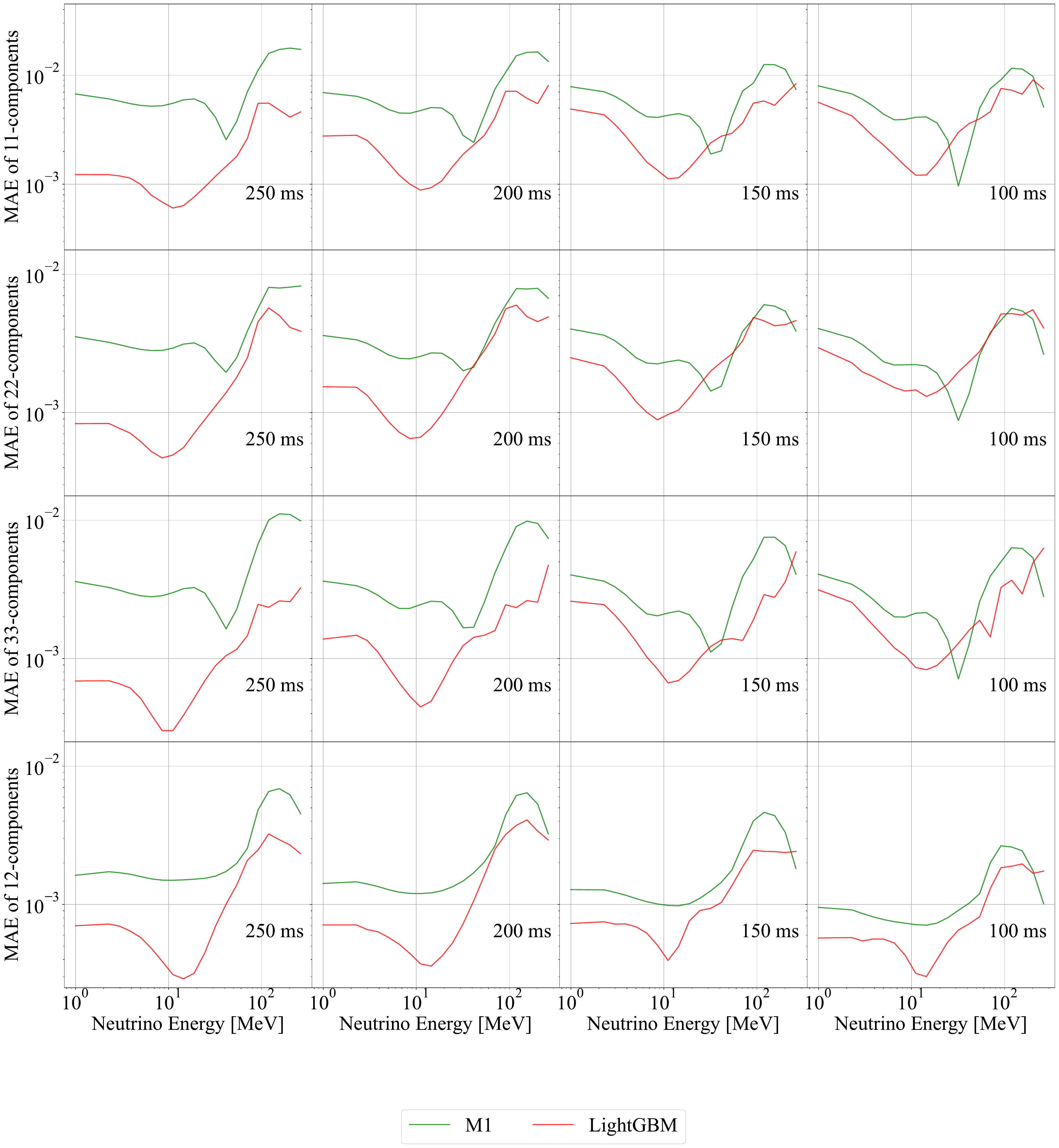}
    \caption{Relationship between neutrino energy and prediction error. Each row corresponds to the plots of \( k_{\rm{FP}}^{11} \), \( k_{\rm{FP}}^{22} \), \( k_{\rm{FP}}^{33} \), and \( k_{\rm{FP}}^{12} \) from top to bottom, while each column represents the plots for 250 ms, 200 ms, 150 ms, and 100 ms after the core bounce from left to right. The green and red lines represent the MAE obtained using the M1 closure relation and our LightGBM prediction, respectively.}
    \label{fig:mae_all}
\end{figure*}

We move on to the predictions of these models. In Figure \ref{fig:mae_all} we plot the mean absolute error (MAE) for the diagonal components of the Eddington tensor for different post-bounce times: 250, 200, 150, and 100 ms. MAE is defined as
\begin{equation}
L_{\varepsilon}^{(1)} = \frac{1}{|D_\varepsilon|} \sum_{D_\varepsilon} |k_{\rm boltz, FP}^{ij} - k_{\rm infer, FP}^{ij}|,
\label{eq:mae_loss}
\end{equation}
where $k_{\rm infer, FP}^{ij}$ is the $ij$-component of the Eddington tensor obtained either with our LightGBM model or via the M1 closure relation. It is apparent that our LightGBM closure gives more accurate predictions than the M1 closure for almost all energies at these times. The behavior of MAE is essentially the same among the three diagonal components. As the time goes back to earlier times, the accuracy is degraded somewhat as expected. It is encouraging, however, that the results are still better than those for the M1 closure except at $\sim$ 30 MeV, where the M1 closure gives exceptionally good results.

\begin{figure*}
    \centering
    \includegraphics[width=1\linewidth]{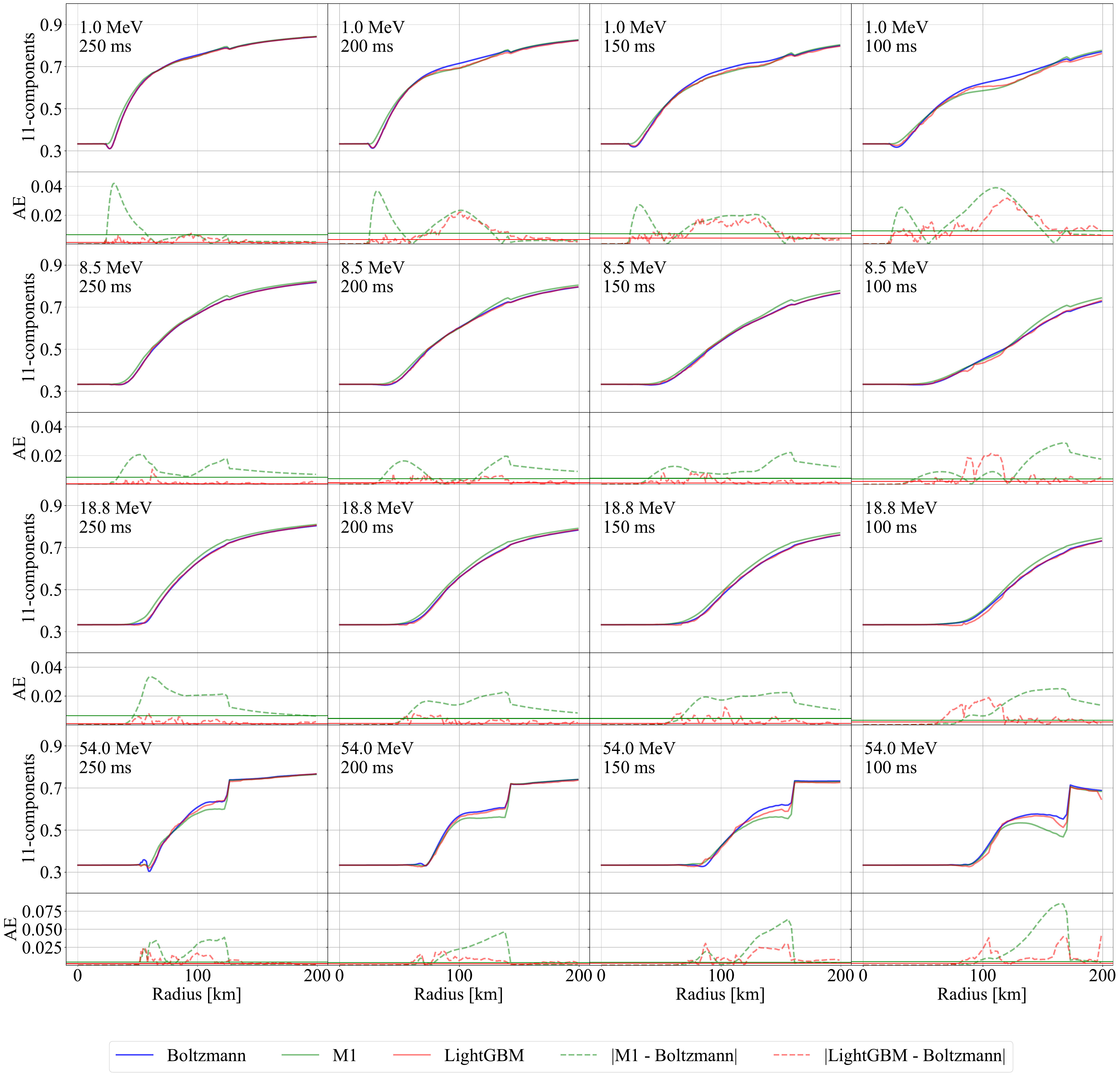}
    \caption{Radial profiles of the $k_{11}$ component of the Eddington tensor at $\theta = \pi/2$. The horizontal axis represents the radius, and the vertical axis shows the $k_{11}$ values. The blue, green, and red lines correspond to the Boltzmann-radiation-hydrodynamics simulations, the M1 closure relation, and the prediction obtained by LightGBM, respectively. The bottom panel shows the absolute values of the differences between the M1 and Boltzmann predictions as $|\rm{M1} - \text{Boltzmann}|$ and those between the LightGBM and Boltzmann predictions as $|\text{LightGBM} - \text{Boltzmann}|$, with solid lines indicating the mean errors.}
    \label{fig:plot_11}
\end{figure*}

\begin{figure*}
    \centering
    \includegraphics[width=1\linewidth]{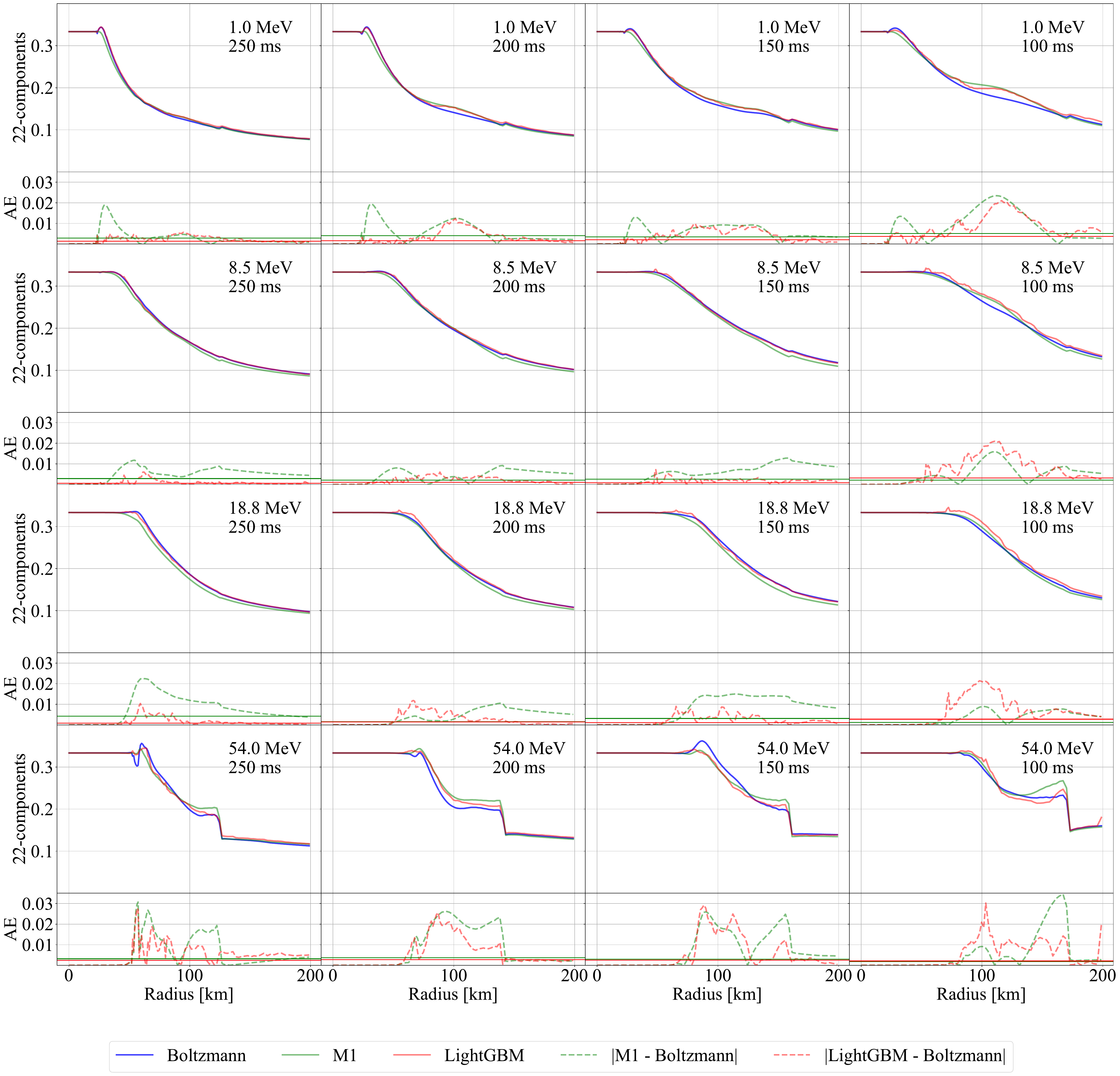}
    \caption{The same as figure \ref{fig:plot_11} except that the target is $k_{\rm{FP}}^{22}$.}
    \label{fig:plot_22}
\end{figure*}

\begin{figure*}
    \centering
    \includegraphics[width=1\linewidth]{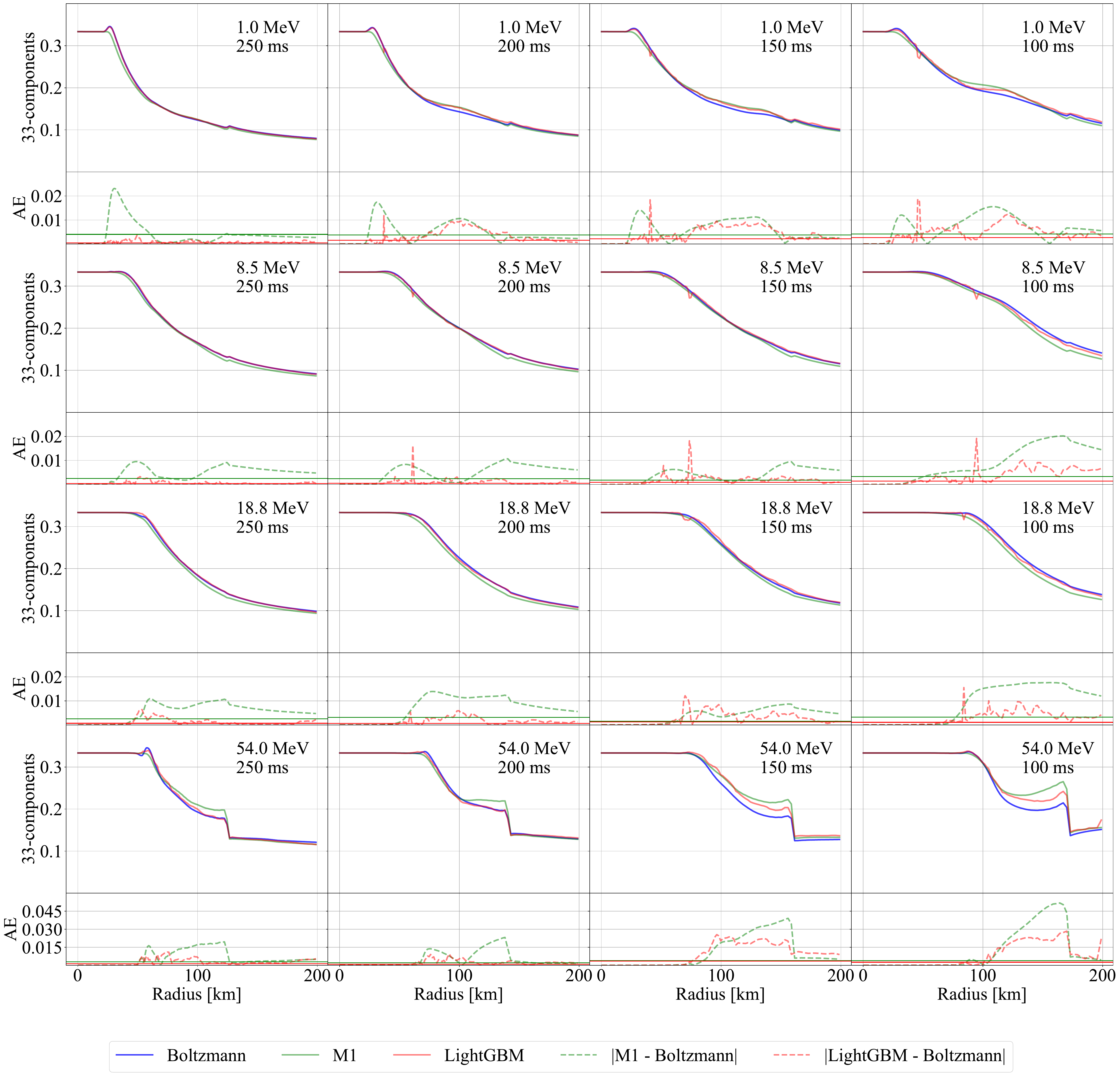}
    \caption{The same as figure \ref{fig:plot_11} except that the target is $k_{\rm{FP}}^{33}$.}
    \label{fig:plot_33}
\end{figure*}

Although MAE is useful to see the overall convergence of the model to the target, what is more important is to what extent the individual components of the Eddington tensor are reproduced at different points in space. We hence compare next the radial profiles of the components of the Eddington tensors along the radial line at $\theta=\pi / 2$ in figures \ref{fig:plot_11}--\ref{fig:plot_33}. The results are not much different for other angles. At the bottom of each plot, we show the absolute error, i.e., the magnitude of deviation from the result of the Boltzmann simulation.

\begin{figure*}
    \centering
    \includegraphics[width=1\linewidth]{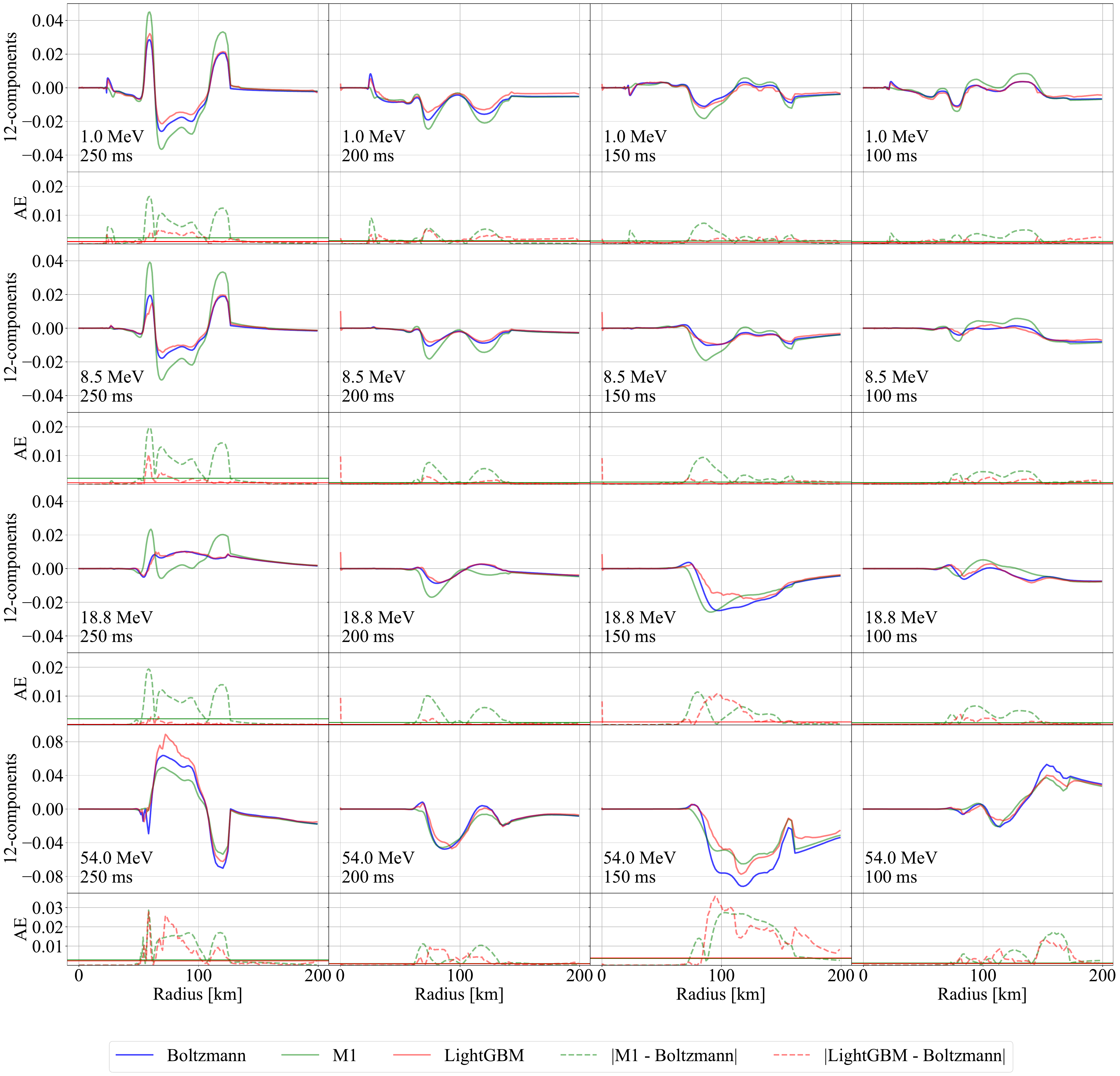}
    \caption{The same as figure \ref{fig:plot_11} except that the target is $k_{\rm{FP}}^{12}$.}
    \label{fig:plot_12}
\end{figure*}

We can confirm again that the LightGBM closure outperforms the M1 closure in general. Of particular note is that our LightGBM model can reproduce the values of the 11-component of the Eddington tensor smaller than 1/3 and the 22- and 33-components greater than 1/3 at rather small radii. Those values are obtained when the opacity decreases rapidly with the radius in the vicinity of the neutrinosphere; then the neutrino distribution is almost isotropic in the outward hemisphere whereas it is heavily depleted in the opposite hemisphere \citep[see the discussion in][]{harada2022deep}. By its construction, the M1 closure cannot treat such angular distributions properly. It is noted that our LightGBM fares even better than the TBNN model in \cite{harada2022deep}.

It is also worth mentioning that the LightGBM model can handle the shock wave with reasonable accuracy. The shock wave is located at $170\,{\rm km}$, $140\,{\rm km}$, $130\,{\rm km}$, $120\,{\rm km}$, and $110\,{\rm km}$ at $100\,{\rm ms}$, $150\,{\rm ms}$, $200\,{\rm ms}$, $250\,{\rm ms}$, and $300\,{\rm ms}$ post-bounce, respectively. The Eddington tensor is discontinuous at the shock wave. The jump gets larger as the energy increases. The radial profiles of the Eddington tensor for the energy of $54\,{\rm MeV}$ presented in Figures \ref{fig:plot_11}--\ref{fig:plot_33} demonstrate indeed that the predictions of the LightGBM model are closer to the results of the Boltzmann simulations than the M1 closure relation in general. 

\added{In order to find out which features play the most influential role in the prediction, we look into the built-in metrics for the feature importance provided by LightGBM. They are given in Table \ref{tab:feature_importance}. The details on the feature importance metrics of LightGBM are given in Appendix ~\ref{appendix:feature_importance}. It is apparent at a glance that the flux factor plays a very significant role as expected. In fact, not only the flux factor itself but also the non-local features derived from it exhibit high feature importance. It is interesting that the collinearity parameter in addition to the baryon density comes next. This indicates that the local information on matter density and motions is important in the accurate prediction of the Eddington tensor.}

It should be noted, however, that the feature importance represents the degree to which the elimination of a single feature degrades the accuracy of prediction. The metric presented here may not fully capture the importance of each feature. It may not reflect its true importance even qualitatively when multiple features work collectively in an intricate way. We hence need caution to interpret the result.

\begin{table}[h!]
\centering
\begin{tabular}{ll}
\hline
\textbf{Features} & \textbf{Feature Importance} \\ \hline
$\chi(r_i, \theta_j)$ & $4.5\times10^5$ \\ 
$\log \chi(r_i, \theta_j)$ & $3.0\times10^5$ \\ 
$\chi(r_{i+1}, \theta_j)$ & $3.5\times10^4$ \\ 
$\log \rho_{\mathrm{baryon}}(r_i, \theta_j)$ & $8.7\times10^3$ \\ 
$\chi(r_{i-5}, \theta_j)$ & $7.0\times10^3$ \\ 
$\chi(r_{i-1}, \theta_j)$ & $6.9\times10^3$ \\ 
$\frac{F^k v_k}{E |v|}(r_{i-10}, \theta_j)$ & $2.5\times10^3$ \\ 
$\epsilon_\nu$ & $2.0\times10^3$ \\ 
$\frac{F^k v_k}{E |v|}(r_{i-5}, \theta_j)$ & $9.8\times10^2$ \\ 
$\frac{F^k v_k}{E |v|}(r_{i}, \theta_j)$ & $9.3\times10^2$ \\ 
$\log \tau(r_{i}, \theta_{j})$ & $9.1\times10^2$ \\ 
$\frac{F^k v_k}{E |v|}(r_{i-1}, \theta_{j})$ & $8.4\times10^2$ \\ 
$\frac{\Delta \chi}{\Delta r}(r_{i}, \theta_{j})$ & $7.0\times10^2$ \\ 
$\chi(r_{i-10}, \theta_{j})$ & $6.1\times10^2$ \\ 
$\chi(r_{i-5}, \theta_{j})$ & $5.3\times10^2$ \\ 
$\frac{F^k v_k}{E |v|}(r_{i+1}, \theta_j)$ & $4.2\times10^2$ \\ 
$\chi(r_{i-30}, \theta_{j})$ & $3.9\times10^2$ \\ 
$\chi(r_{i-1}, \theta_{j})$ & $3.4\times10^2$ \\ 
$\chi(r_{i-15}, \theta_{j})$ & $2.1\times10^2$ \\ 
$\frac{F^k v_k}{E |v|}(r_{i-15}, \theta_j)$ & $1.6\times10^2$ \\ 
$\chi(r_{i}, \theta_{j+1})$ & $1.6\times10^2$ \\ 
$\chi(r_{i-25}, \theta_{j})$ & $1.5\times10^2$ \\ 
$\Delta \Bigl(\frac{F^k v_k}{E |v|}\Bigr)/\Delta \theta(r_{i}, \theta_{j})$ & $1.5\times10^2$ \\ 
$\chi(r_{i-20}, \theta_{j})$ & $1.3\times10^2$ \\ 
$\frac{F^k v_k}{E |v|}(r_{i-25}, \theta_{j})$ & $1.2\times10^2$ \\ 
$\frac{F^k v_k}{E |v|}(r_{i+1}, \theta_{j})$ & $1.2\times10^2$ \\ 
$\Delta \Bigl(\frac{F^k v_k}{E |v|}\Bigr)/\Delta r(r_{i}, \theta_{j})$ & $9.9\times10^1$ \\ 
$\frac{F^k v_k}{E |v|}(r_{i+5}, \theta_{j})$ & $9.1\times10^1$ \\ 
$\frac{F^k v_k}{E |v|}(r_{i-1}, \theta_{j})$ & $7.4\times10^1$ \\ 
$\frac{\Delta \chi}{\Delta \theta}(r_{i}, \theta_{j})$ & $7.1\times10^1$ \\ 
$\frac{F^k v_k}{E |v|}(r_{i-20}, \theta_{j})$ & $6.9\times10^1$ \\ 
$\frac{F^k v_k}{E |v|}(r_{i-30}, \theta_{j})$ & $6.8\times10^1$ \\ 
$I(\tau \geq 5)(r_{i}, \theta_{j})$ & $2.3\times10^1$ \\ 
$\text{sign}(v^2)(r_{i}, \theta_{j})$ & $2.0\times10^1$ \\ 
$\text{sign}(v^1)(r_{i}, \theta_{j})$ & $3.2\times10^{-1}$ \\ \hline
\end{tabular}
\caption{Feature importance for each feature in the model predicting the 11-component of the Eddington tensor (values rounded to 2 significant figures). The subscript LB for the flux factor and the collinearity parameter are omitted for simplicity.}
\label{tab:feature_importance}
\end{table}

The Eddington tensor predicted by LightGBM has several issues. Firstly, the predicted values are not as smooth as those obtained with the M1 closure. This tendency becomes more pronounced as we go back in time and see the predictions themselves degraded. In fact, the Eddington tensor at 250 ms post-bounce plotted in Figures \ref{fig:plot_11}--\ref{fig:plot_33} is relatively smooth, whereas the lines are more jagged at 100 ms after bounce as shown in the same figures. We think that this phenomenon occurs because the current predictions are made for each grid point individually and that the issue could be resolved by adopting a machine learning model capable of incorporating non-local information into the predictions. Machine learning models such as Convolutional Neural Networks (CNNs) \citep{lecun1998gradient} or Long Short Term Memory (LSTM) \citep{10.1162/neco.1997.9.8.1735} may have robustness to local noises and outliers in the input data, thereby giving smoother predictions. \added{It is well known that the use of simple neural networks in image generation tasks often results in outputs with significant noises and blurred edges. This is thought to occur because they give at individual points predictions that do not take into account global contextual information just as our current model does not (see, e.g., \citep{foster2019generative}). By contrast, CNNs, which reflect such information, can produce cleaner and more coherent images in the predictions at different points. We expect the same for the current task by adopting CNNs or LSTMs.}

\begin{figure*}
    \centering
    \includegraphics[width=0.8\linewidth]{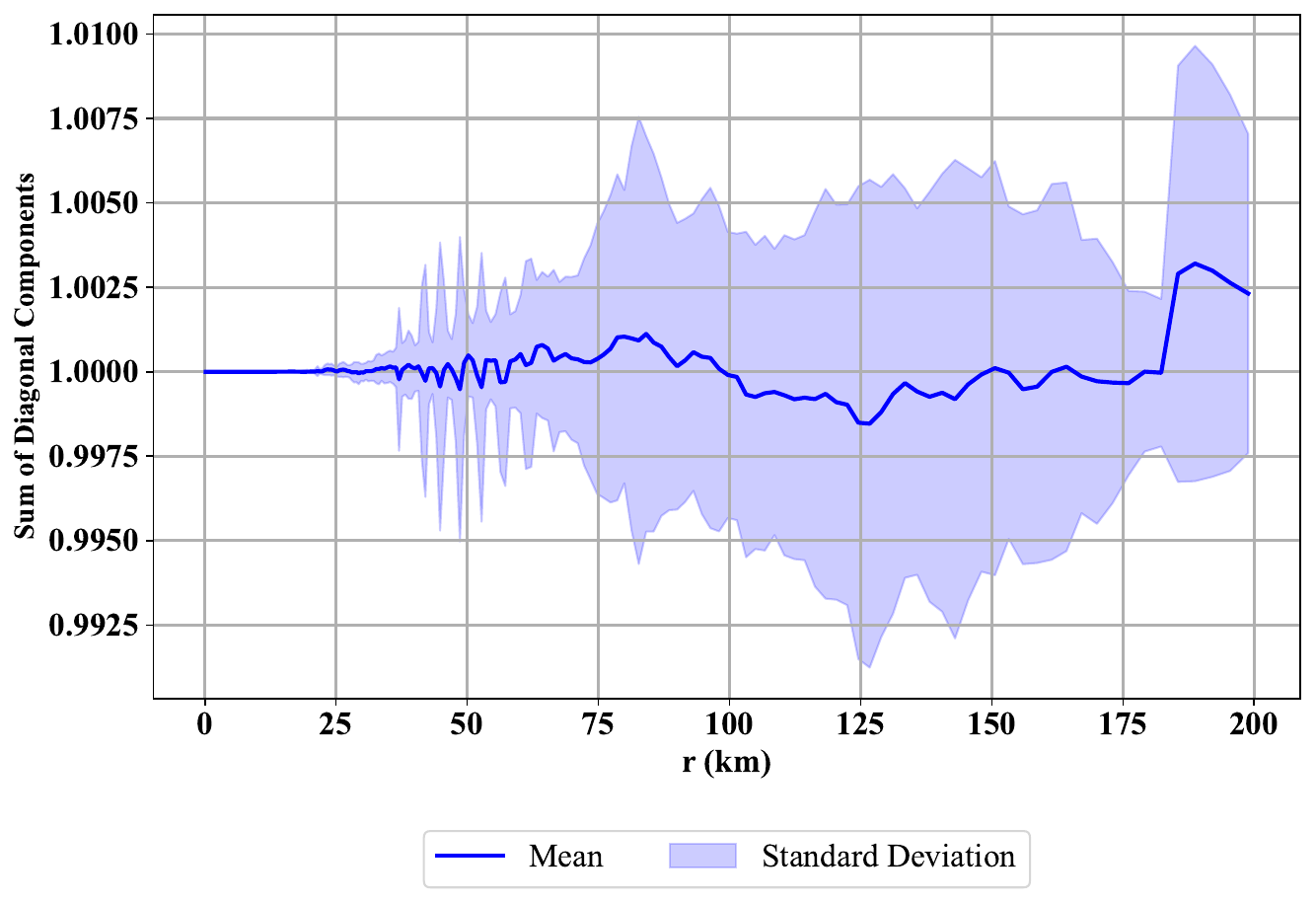}
    \caption{Mean and standard deviation of the sum of diagonal components over the zenith angle $\theta$ and the neutrino energy $\varepsilon_{\nu}$ as a function of radius at 200 ms after core bounce.}
    \label{fig:e_conv}
\end{figure*}

It should be also mentioned that the current model does not take into account physical requirements. For instance, the trace of the Eddington tensor is unity by definition. There is no guarantee, however, that the predicted results satisfy this constraint although the violation is quite minor in the current models, as can be seen in Figure \ref{fig:e_conv}. This issue may be addressed with physics-informed neural networks \citep{raissi2019physics}: for example, one may incorporate the violations of constraints into the loss function so that they could be minimized. \added{However, LightGBM is not suited for this physics-informed scheme at least for the current version.} Although these possibilities are interesting in their own right and warrant further investigation, they are much beyond the scope of this paper and will be addressed in future work.

\subsection{Off-Diagonal Component}
Although the diagonal components, particularly the 11-component, are dominant in neutrino transport, the off-diagonal components should not be forgotten \citep[see discussions in][]{2018ApJ...854..136N}. Designed to reproduce the dominant component, the M1 closure is known to give large errors to the off-diagonal components sometimes \citep{2019ApJ...872..181H,2020ApJ...902..150H}. It is hence one of our goals in the machine learning modeling of the Eddington tensor to better reproduce the off-diagonal components. It is admittedly true in this respect that the previous TBNN closure was not so good, either \citep{harada2022deep}. In the following, we examine the accuracy of the off-diagonal components in our LightGBM model.
 
We focus on the 12-component, which is actually the most important off-diagonal component. Figure \ref{fig:mae_all} shows the MAE for this component as a function of the neutrino energy at different post-bounce times. We find again that our LightGBM model achieves higher accuracies in general compared to the M1 closure. The radial profiles of the $12$-component of the Eddington tensor are presented in Figure \ref{fig:plot_12}. They are obtained for the radial ray at $\theta=\pi/2$. At 250 ms post-bounce, our LightGBM model fares better than the M1 closure except at $\varepsilon_{\nu}$ = 54.0MeV, where the two models give similar errors. As we go back in time, the deviation from the results of the Boltzmann simulation gets larger for the LightGBM model and its advantage becomes not so remarkable.

A potential factor that contributes to the compromised success is the variation in the statistical properties among the data at different time steps. Tables \ref{tab:statistics-11} and 
\ref{tab:statistics-12} give the mean and standard deviation of $k_{\rm boltz, FP}^{11}$ and $k_{\rm boltz, FP}^{12}$ at each time step. It is evident that the statistical properties of $k_{\rm boltz, FP}^{12}$ are significantly different from time to time compared to those of $k_{\rm boltz, FP}^{11}$. It may be that the generalization should not have been expected for the  off-diagonal component in the first place. Standardisation of the dataset across all time steps may be a solution but again is out of the scope of this paper.

\deleted{It is worth noting that different features were utilized for the diagonal and off-diagonal components in our LightGBM model. Specifically for the off-diagonal components, it was unclear which features held the most significance. In predicting the diagonal components, we utilized 35 features, whereas for the off-diagonal components, 317 features were employed. For example, features such as fluid velocity, which were removed when predicting diagonal components, are included. Off-diagonal components are more significantly affected by factors such as fluid motion compared to diagonal components, which suggests that background information may be more important for their prediction.}

\begin{table}[h]
    \centering
    \begin{tabular}{cccc}
        \hline
        Timestep (ms) & Mean & Standard deviation \\
        \hline
        100 & 0.38 & 0.10 \\
        150 & 0.39 & 0.12 \\
        200 & 0.40 & 0.13 \\
        250 & 0.42 & 0.15 \\
        300 & 0.43 & 0.16 \\
        \hline
    \end{tabular}
    \caption{Statistics of $k_{boltz, \rm{FP}}^{11}$ at Different Timesteps}
    \label{tab:statistics-11}
\end{table}

\begin{table}[h]
    \centering
    \begin{tabular}{cccc}
        \hline
        Timestep (ms) & Mean & Standard deviation \\
        \hline
        100 & $5.54 \times 10^{-4}$ & $1.01 \times 10^{-2}$ \\
        150 & $1.30 \times 10^{-4}$ & $1.41 \times 10^{-2}$ \\
        200 & $2.29 \times 10^{-4}$ & $1.59 \times 10^{-2}$ \\
        250 & $1.79 \times 10^{-3}$ & $1.61 \times 10^{-2}$ \\
        300 & $9.48 \times 10^{-4}$ & $1.60 \times 10^{-2}$ \\
        \hline
    \end{tabular}
    \caption{Statistics of $k_{boltz, \rm{FP}}^{12}$ at Different Timesteps}
    \label{tab:statistics-12}
\end{table}

\subsection{Importance of feature engineering}
Feature engineering plays a crucial role in predicting the Eddington tensor accurately. In this work, we incorporate many \replaced{advanced features}{combined features} indeed as listed in Table \replaced{\ref{tab:Advanced_features}}{\ref{tab:features}}. It is possible because LightGBM is very efficient particularly in training, ignoring irrelevant features automatically. As explained in section 3.2, we find that the flux factor and the \replaced{spatial-shift}{non-local} are the two most important features for the improvement of accuracy, which we demonstrate here. 

\begin{figure*}
    \centering
    \includegraphics[width=0.7\linewidth]{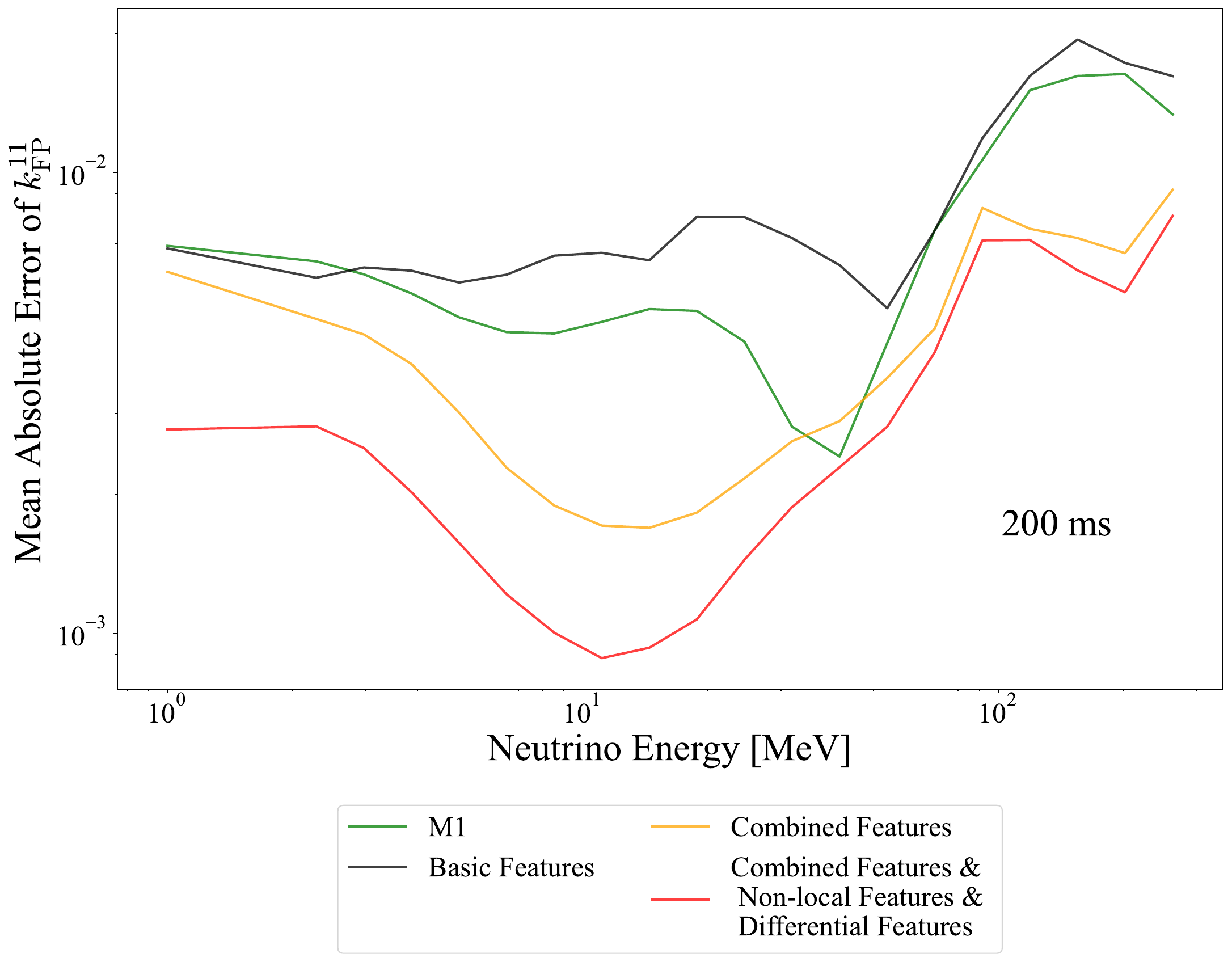}
    \caption{The same as the bottom left figure in figure \ref{fig:mae_all} except there are some additional plots: the black one is the LightGBM prediction with basic features \replaced{(Table \ref{tab:basic_feat})}{(ll. 1--6 of Table \ref{tab:features})} and the orange line with \replaced{advanced features}{combined features}.}
    \label{fig:feateng_mae}
\end{figure*}

\begin{figure*}
    \centering
    \includegraphics[width=0.8\linewidth]{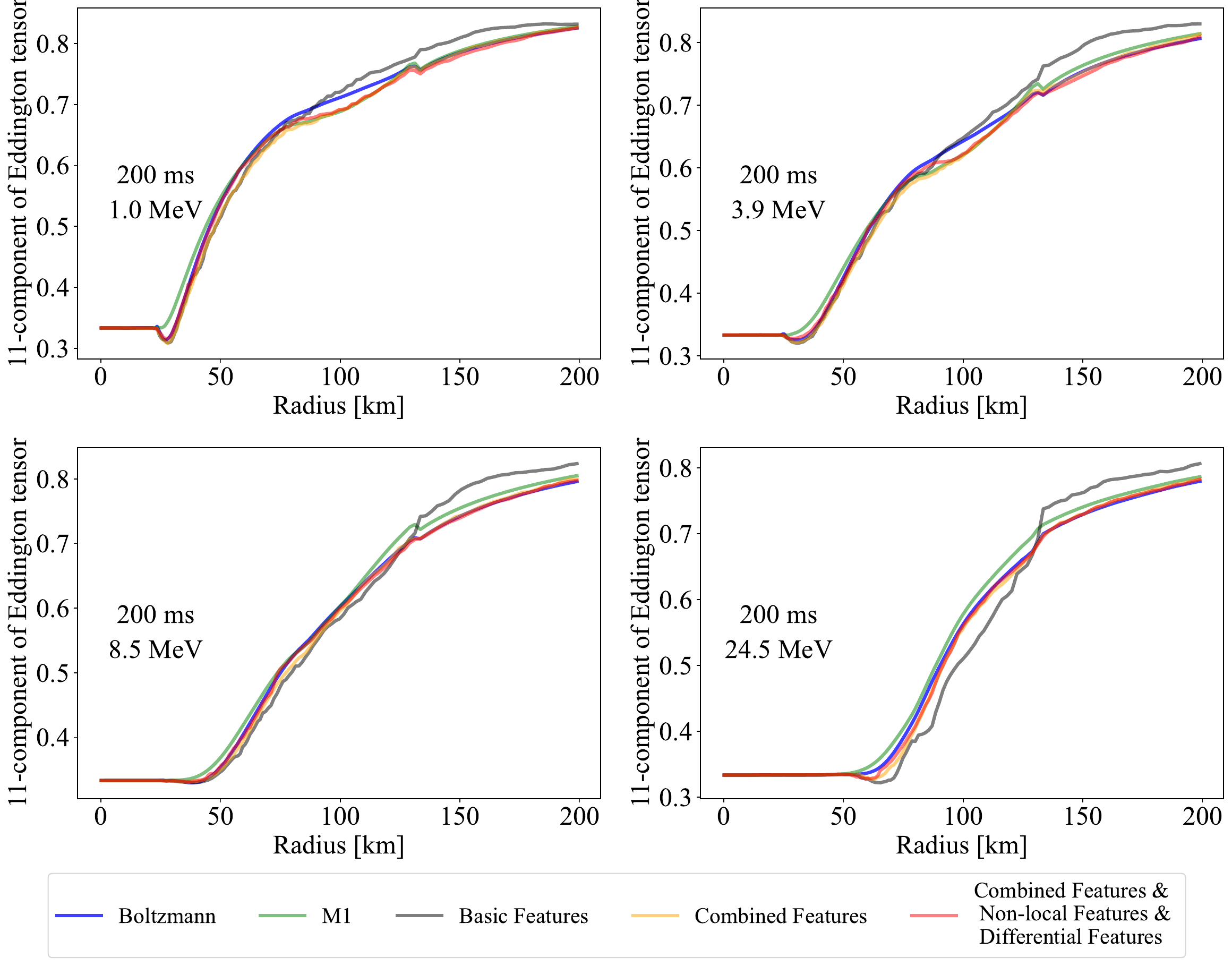}
    \caption{LightGBM prediction of $k_{\rm{FP}}^{11}$ with different features $(200\,{\rm ms})$. The blue line corresponds to the Boltzmann-radiation-hydrodynamics code and the green line to the M1 closure method. The black line is the LightGBM prediction with basic features \replaced{(Table \ref{tab:basic_feat})}{(ll. 1--6 of Table \ref{tab:features})}, the orange line with \replaced{advanced features}{combined features}, and the red line with all engineered features: \replaced{advanced features}{combined features}, \replaced{spatial-shift}{non-local} features and \replaced{spatial-difference features}{differential features}.}
    \label{fig:feateng_plot}
\end{figure*}

In Figure \ref{fig:feateng_mae}, we show how MAE gets better from the one for the model with only the basic features included. The MAE for the model with the basic features alone is actually larger than that for the M1 closure method for most of the neutrino energy. The MAE becomes smaller, however, once the \replaced{advanced features}{combined features} are incorporated, particularly for energies close to the mean energy. It is further improved if the \replaced{spatial-shift}{non-local} and \replaced{difference features}{differential features} are taken into account in addition. The same trend is observed for the Eddington tensor $k_{\rm{FP}}^{11}$ itself as a function of radius as demonstrated in Figure \ref{fig:feateng_plot}.

What we learned from these experiments on feature engineering may be three-fold. Firstly, the local low-order moments, the inputs for the ordinary closure methods, are not sufficient to reproduce the Eddington tensor accurately. We find that the information on the background matter such as the baryon density or the optical depth is particularly important. Secondly, the non-local information on the neutrino distribution is probably the most crucial. Although their incorporation in LightGBM is rather straightforward, it will not be so easy for the analytic closure relation to take them into account. Thirdly, finding nice combinations of the basic features is also an important thing. Our LightGBM model is useful in this respect since we can try many possible combinations in a relatively short time. The new features so obtained may be employed in other machine learning models.

\begin{figure}
    \centering
    \includegraphics[width=1\linewidth]{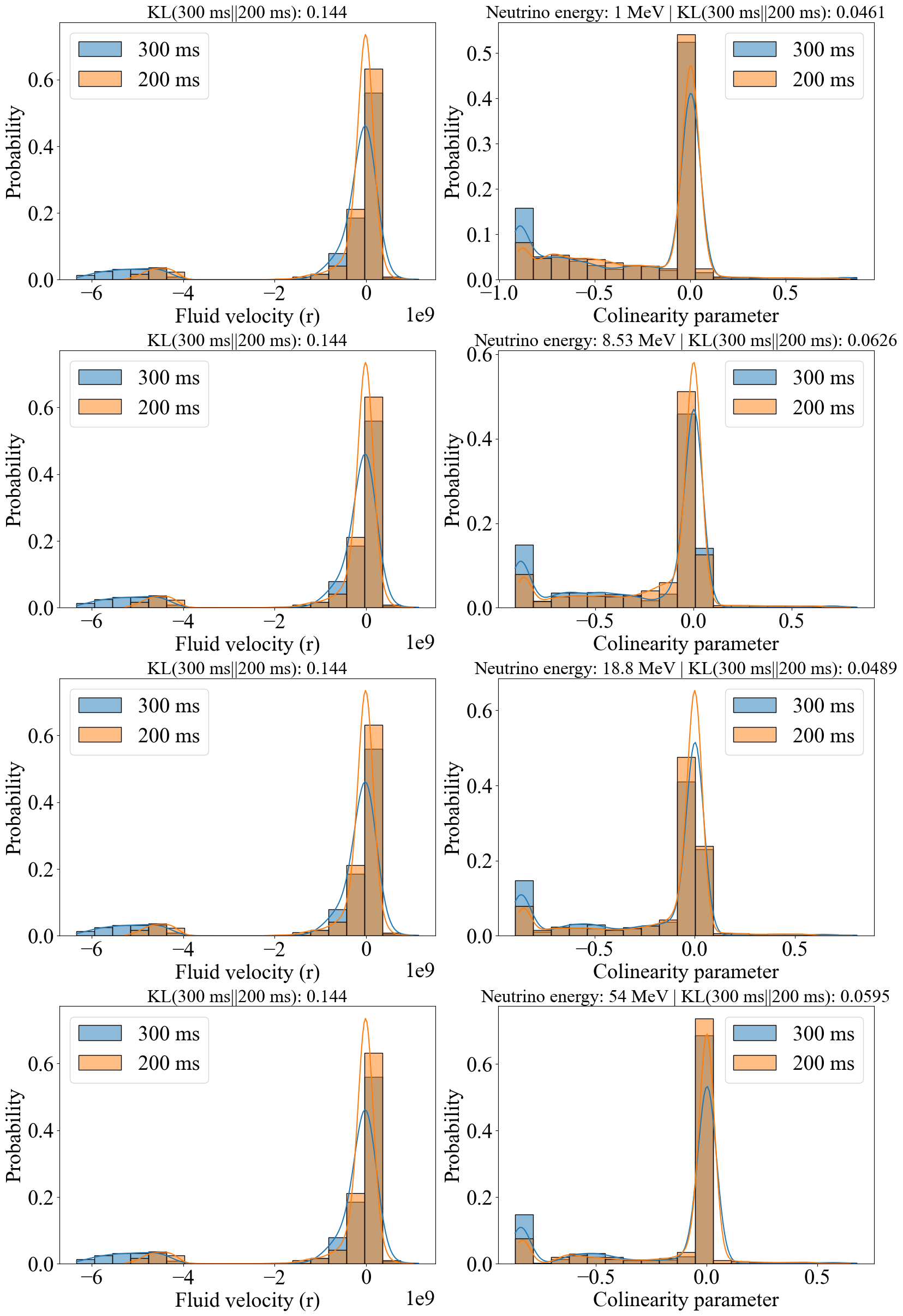}
    \caption{Probability distributions of fluid velocity (”Fluid velocity (r)”) and the collinearity parameter(”collinearity parameter”) for neutrinos with the energy of 1.0 MeV, 8.5 MeV, 18.8 MeV, and 54.0 MeV. Each row corresponds to one of these neutrino energies. The data sets at 300 ms and 200 ms post-bounce are colored blue and orange, respectively. The histograms and KDE (Kernel Density Estimation) plots illustrate the statistical difference between these two data sets for each neutrino energy level. The value of the KL-divergence is given at the plots to quantify the extent of the differences.}
    \label{fig:stat_analysis}
\end{figure}

\added{Finally, from a statistical perspective, feature engineering may be understood as a process that standardizes the distributions of inputs from different snapshots. Indeed, as shown in Figure \ref{fig:stat_analysis}, the collinearity parameter generated through feature engineering exhibits smaller differences among snapshots compared to the radial component of fluid velocity in the basic features. This observation is quantitatively supported by the Kullback-Leibler (KL) divergence, defined as
\[
D_{\mathrm{KL}}(P \| Q) = \int_{-\infty}^\infty P(x) \log\frac{P(x)}{Q(x)} \, dx,
\]
where \(P(x)\) and \(Q(x)\) represent the probability density functions of the distributions from two different snapshots. The KL divergence, introduced by \cite{10.1214/aoms/1177729694}, measures the difference between two probability distributions; the smaller the value is, the more similar the two distributions are to each other. The lower KL divergence obtained for the collinearity parameter indicates that feature engineering is indeed useful in reducing the difference between the distributions taken from different snapshots.}

\section{Conclusion \& Discussion}
In this paper, we develop a machine learning model using LightGBM to predict the Eddington tensor, or the second moment, of the angular distribution in momentum space of neutrinos often employed in the core-collapse supernova simulation. Unlike the ordinary closure relations, we employ not only low-order moments of the neutrino angular distribution in momentum space but also exploit the information on the matter distribution as well as non-local features of the neutrino distribution. For the training and validation of the machine, we utilize the numerical result of the Boltzmann-radiation-hydrodynamics simulations of the core-collapse of a non-rotating 15$M_{\odot}$ progenitor model. The training data are a snapshot at 300 ms after core bounce, while the validation data are snapshots taken at 100 ms, 150 ms, 200 ms, and 250 ms post-bounce. For all these times, our machine learning model shows better accuracy than the M1 closure in predicting the Eddington tensor, highlighting the potential of machine learning to be an alternative closure relation.

The key factor in the model building is the feature engineering process. It is true that finding features of the greatest relevance in improving the prediction is the most critical but it is equally important to produce nice combinations of the basic features that the machine can easily understand. Those features should have some physical meanings so that they could be applied to different evolutionary stages. The LightGBM, a variant of the Gradient Boosting Decision Tree, is efficient in the management of memory and time, allowing us to handle a large number of features at a time quickly and, as a result, identify the features that improve generalization ability substantially. In fact, LightGBM is able to reproduce better both diagonal and off-diagonal components of the Eddington tensor in most cases than the M1 closure although the accuracy degrades somewhat as we go back to earlier post-bounce times. In particular, the dip in the 11-component near the neutrino sphere is represented quite well, which not only the M1 closure but our previous tensor-basis neural network model is unable to do. The hump in the 12-component in the same region is represented reasonably as well.

\added{
The ultimate goal is to implement a machine learning model in a CCSN simulation code with the truncated moment method and run it for realistic simulations. Before doing so, however, several critical issues should be addressed.

The first issue is the inference time. Although LightGBM offers advantages such as short training times and the ability to handle numerous features, it has limitations in GPU-based parallel computation during prediction. The current model is not fast enough for real-time predictions in simulations. The second issue pertains to the generalization performance. While the predictive accuracy of the current model has significantly improved from the previous work, its generalization performance is degraded as the temporal separation between the training and testing data increases. Moreover, the ability to generalize to different progenitors remains to be studied: a model trained on the data from one progenitor should be applicable to other stars with different masses (and possibly rotations). This study is limited to the generalization within the same progenitor. The third issue concerns the imposition of physical/mathematical constraints. The trace of the Eddington tensor, for example, is equal to unity by definition. As shown in Figure \ref{fig:e_conv}, however, a small error persists in the current model. Reducing such errors as much as possible is necessary to achieve more precise simulations.

To address these challenges, we do not think that the LightGBM model is the best choice and we are building other closure models based on the neural network. While previous studies have employed neural network-based models, our models had limited generalization capabilities. We think that the employment of the extended features that are judged useful in this study in neural networks will improve the performance substantially. In fact, neural networks will offer shorter inference times and greater flexibility  in imposing constraints. Some neural networks \citep{lecun1998gradient, 10.1162/neco.1997.9.8.1735} can reflect non-local information in the point-wise prediction. With a proper combination of a neural network and the extended features, we will hopefully be able to address all three issues mentioned above.
}

\section*{Acknowledgments}

We acknowledge Hiroki Nagakura and Keiya Hirashima for fruitful discussions. 
This work was supported by KAKENHI Grant Numbers JP21K13913, 21H01083.
This work was also supported by MEXT as ``Program for Promoting Researches on the Supercomputer Fugaku'' (Toward a unified view of the universe: from large scale structures to planets).
S. T. is supported by International Graduate Program of Innovation for Intelligent World.
S. Y. is supported by Institute for Advanced Theoretical and Experimental Physics, Waseda University and the Waseda University Grant for Special Research Projects (project number: 2023-C141, 2024-C56, 2024-Q014).
We acknowledge the high-performance computing resources of the K-computer / the supercomputer Fugaku provided by RIKEN, the FX10 provided by Tokyo University, the FX100 provided by Nagoya University, the Grand Chariot provided by Hokkaido University, and Oakforest-PACS / Wisteria Odyssey provided by JCAHPC through the HPCI System Research Project (Project ID: hp130025, hp230204, hp240219, hp240264, 140211, 150225, 150262, 160071, 160211, 170031, 170230, 170304, 180111, 180179, 180239, 190100, 190160, 200102, 200124, 210050, 220047, 220223, 230056, 230270, 240041) for producing and processing the supervisor data.

\software{}
LightGBM (\cite{NIPS2017_6449f44a})

\bibliography{ref}{}
\bibliographystyle{aasjournal}

\added{
\appendix
\section{Appendix: Derivation of Feature Importance (Gain) in LightGBM}
\label{appendix:feature_importance}
Feature importance in LightGBM is often measured by \textit{gain}, which quantifies the improvement in the objective function due to splits involving a specific feature. This section outlines the derivation of gain. In a decision tree, a split at a node is performed to minimize the loss function. The improvement in the loss function, referred to as the \textit{gain}, is calculated as:
\[
\text{Gain}(s) = L_{\text{parent}} - (L_{\text{left}} + L_{\text{right}})
\]
where \( L_{\text{parent}} \), \( L_{\text{left}} \), and \( L_{\text{right}} \) are the losses at the parent node, left child node, and right child node, respectively. LightGBM uses gradient statistics to approximate these losses. For a given node, the first-order gradient \( G \) and second-order gradient \( H \) are computed as:
\[
G = \sum_{i \in \text{node}} g_i, \quad H = \sum_{i \in \text{node}} h_i
\]
where \( g_i \) and \( h_i \) are the first and second derivatives of the loss function for each data point \( i \). The gain for a split is then approximated as:
\[
\text{Gain}(s) = \frac{G_{\text{left}}^2}{H_{\text{left}} + \lambda} + \frac{G_{\text{right}}^2}{H_{\text{right}} + \lambda} - \frac{G_{\text{parent}}^2}{H_{\text{parent}} + \lambda}
\]
Here, \( G_{\text{parent}}, H_{\text{parent}} \), \( G_{\text{left}}, H_{\text{left}} \), and \( G_{\text{right}}, H_{\text{right}} \) are the gradient and Hessian sums for the parent, left child, and right child nodes, respectively, and \( \lambda \) is a regularization parameter. The feature importance for a feature \( x_j \) is obtained by summing the gains from all splits where \( x_j \) is used:
\[
\text{Feature Importance (Gain)}_{x_j} = \sum_{s \in \text{splits using } x_j} \text{Gain}(s)
\]
LightGBM computes this metric during training, aggregating the contributions of each feature across all trees in the model. Thus, gain provides an interpretable measure of how much a feature contributes to improving the model's predictive performance.
}

\listofchanges

\end{document}